\documentclass[twoside,twocolumn,9pt]{article}
\usepackage{extsizes}
\usepackage[super,sort&compress,comma]{natbib} 
\usepackage[version=3]{mhchem}
\usepackage[left=1.5cm, right=1.5cm, top=1.785cm, bottom=2.0cm]{geometry}
\usepackage{balance}
\usepackage{times,mathptmx}
\usepackage{sectsty}
\usepackage{graphicx} 
\usepackage{lastpage}
\usepackage[format=plain,justification=justified,singlelinecheck=false,font={stretch=1.125,small,sf},labelfont=bf,labelsep=space]{caption}
\usepackage{float}
\usepackage{fancyhdr}
\usepackage{fnpos}
\usepackage[english]{babel}
\usepackage{array}
\usepackage{droidsans}
\usepackage{charter}
\usepackage[T1]{fontenc}
\usepackage[usenames,dvipsnames]{xcolor}
\usepackage{setspace}
\usepackage[compact]{titlesec}
\usepackage{amssymb}

\definecolor{cream}{RGB}{222,217,201}

\begin{document}

\pagestyle{fancy}
\thispagestyle{plain}
\fancypagestyle{plain}{

\fancyhead[C]{\includegraphics[width=18.5cm]{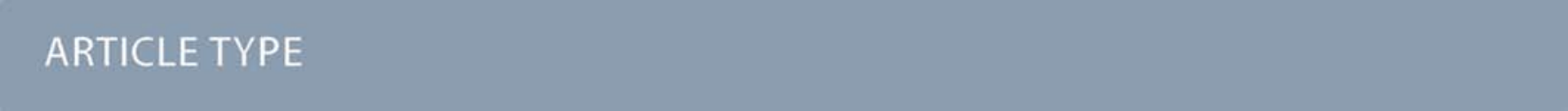}}
\fancyhead[L]{\hspace{0cm}\vspace{1.5cm}\includegraphics[height=30pt]{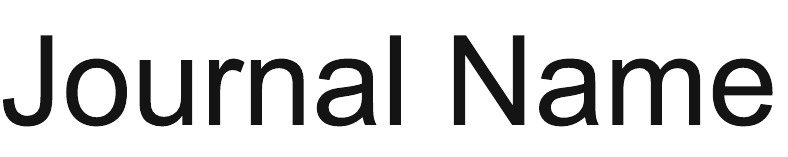}}
\fancyhead[R]{\hspace{0cm}\vspace{1.7cm}\includegraphics[height=55pt]{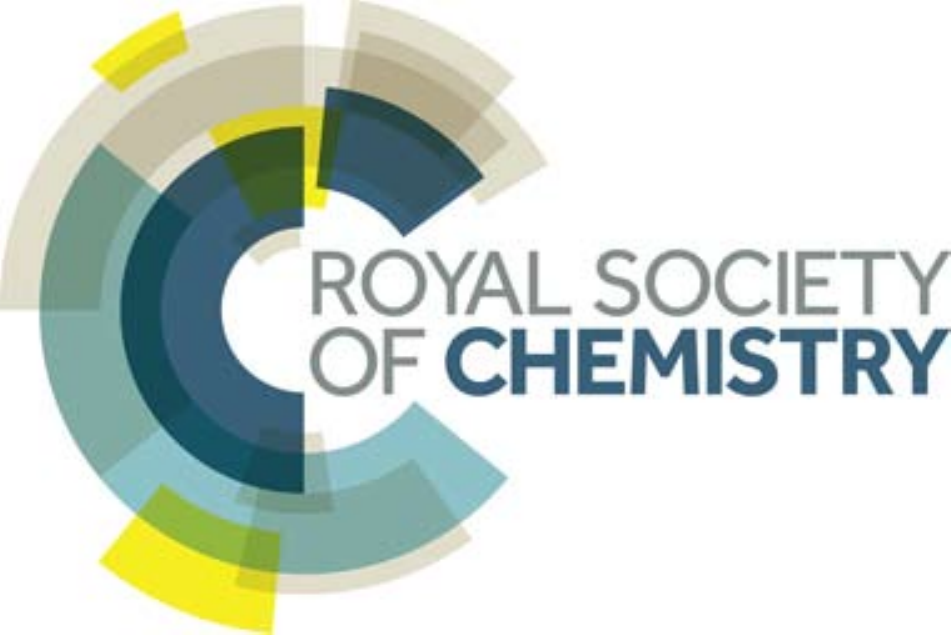}}
\renewcommand{\headrulewidth}{0pt}
}

\makeFNbottom
\makeatletter
\renewcommand\LARGE{\@setfontsize\LARGE{15pt}{17}}
\renewcommand\Large{\@setfontsize\Large{12pt}{14}}
\renewcommand\large{\@setfontsize\large{10pt}{12}}
\renewcommand\footnotesize{\@setfontsize\footnotesize{7pt}{10}}
\makeatother

\renewcommand{\thefootnote}{\fnsymbol{footnote}}
\renewcommand\footnoterule{\vspace*{1pt}%
\color{cream}\hrule width 3.5in height 0.4pt \color{black}\vspace*{5pt}} 
\setcounter{secnumdepth}{5}

\makeatletter 
\renewcommand\@biblabel[1]{#1}            
\renewcommand\@makefntext[1]%
{\noindent\makebox[0pt][r]{\@thefnmark\,}#1}
\makeatother 
\renewcommand{\figurename}{\small{Fig.}~}
\sectionfont{\sffamily\Large}
\subsectionfont{\normalsize}
\subsubsectionfont{\bf}
\setstretch{1.125} 
\setlength{\skip\footins}{0.8cm}
\setlength{\footnotesep}{0.25cm}
\setlength{\jot}{10pt}
\titlespacing*{\section}{0pt}{4pt}{4pt}
\titlespacing*{\subsection}{0pt}{15pt}{1pt}

\fancyfoot{}
\fancyfoot[LO,RE]{\vspace{-7.1pt}\includegraphics[height=9pt]{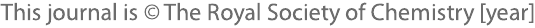}}
\fancyfoot[CO]{\vspace{-7.1pt}\hspace{13.2cm}\includegraphics{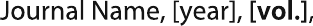}}
\fancyfoot[CE]{\vspace{-7.2pt}\hspace{-14.2cm}\includegraphics{head_foot/RF}}
\fancyfoot[RO]{\footnotesize{\sffamily{1--\pageref{LastPage} ~\textbar  \hspace{2pt}\thepage}}}
\fancyfoot[LE]{\footnotesize{\sffamily{\thepage~\textbar\hspace{3.45cm} 1--\pageref{LastPage}}}}
\fancyhead{}
\renewcommand{\headrulewidth}{0pt} 
\renewcommand{\footrulewidth}{0pt}
\setlength{\arrayrulewidth}{1pt}
\setlength{\columnsep}{6.5mm}
\setlength\bibsep{1pt}

\makeatletter 
\newlength{\figrulesep} 
\setlength{\figrulesep}{0.5\textfloatsep} 

\newcommand{\topfigrule}{\vspace*{-1pt}%
\noindent{\color{cream}\rule[-\figrulesep]{\columnwidth}{1.5pt}} }

\newcommand{\botfigrule}{\vspace*{-2pt}%
\noindent{\color{cream}\rule[\figrulesep]{\columnwidth}{1.5pt}} }

\newcommand{\dblfigrule}{\vspace*{-1pt}%
\noindent{\color{cream}\rule[-\figrulesep]{\textwidth}{1.5pt}} }

\renewcommand\floatpagefraction{.99}
\renewcommand\topfraction{.99}
\renewcommand\bottomfraction{.99}
\renewcommand\textfraction{.01}
\renewcommand\dbltopfraction{0.99}
\renewcommand\dblfloatpagefraction{0.99}

\makeatother

\twocolumn[
  \begin{@twocolumnfalse}
\vspace{3cm}
\sffamily
\begin{tabular}{m{4.5cm} p{13.5cm} }

\includegraphics{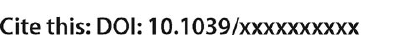} & \noindent\LARGE{\textbf{Columnar Shifts as Symmetry-Breaking Degrees of Freedom in Molecular 
Perovskites}} \\
\vspace{0.3cm} & \vspace{0.3cm} \\

 & \noindent\large{Hanna L. B. Bostr\"om, Joshua A. Hill, and Andrew L. Goodwin$^\ast$} \\
 
\includegraphics{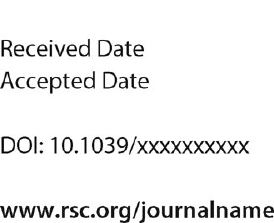} & \noindent\normalsize{We introduce columnar shifts---collective rigid-body translations---as a structural degree of freedom relevant to the phase behaviour of molecular perovskites ABX$_{\textrm3}$ (X = molecular anion). Like the well-known octahedral tilts of conventional perovskites, shifts also preserve the octahedral coordination geometry of the B-site cation in molecular perovskites, and so are predisposed to influencing the low-energy dynamics and displacive phase transitions of these topical systems. We present a qualitative overview of the interplay between shift activation and crystal symmetry breaking, and introduce a generalised terminology to allow characterisation of simple shift distortions, drawing analogy to the ``Glazer notation'' for octahedral tilts. We apply our approach to the interpretation of a representative selection of azide and formate perovskite structures, and discuss the implications for functional exploitation of shift degrees of freedom in negative thermal expansion materials and hybrid ferroelectrics.} \\

\end{tabular}

 \end{@twocolumnfalse} \vspace{0.6cm}

  ]

\renewcommand*\rmdefault{bch}\normalfont\upshape
\rmfamily
\section*{}
\vspace{-1cm}


\footnotetext{\textit{Inorganic Chemistry Laboratory, Department of Chemistry, University of Oxford, South Parks Road, Oxford OX1 3QR, UK. Tel: +44 (0)1865 272137; E-mail: andrew.goodwin@chem.ox.ac.uk.}}

\section{Introduction}

A large variety of important physical properties of perovskite oxides are the result of symmetry-breaking processes that involve ordering of structural, electronic, or magnetic degrees of freedom. From a materials design viewpoint, the role of octahedral tilts is especially important because so-called ``tilt engineering'' approaches now allow control over macroscopic polarisation\cite{Benedek2011} and magnetisation.\cite{Pitcher2015,Benedek2015} Moreover, because tilt degrees of freedom often dominate the vibrational behaviour of perovskites, the same approaches can allow control over dynamical phenomena such as negative thermal expansion (NTE).\cite{Senn2016a} Central to these design approaches is an understanding of the ways in which activation of specific tilt systems can affect space group symmetry.\cite{Glazer1972,Howard1998,Howard2003} Formally, this relationship is given by the irreducible representation of a given tilt distortion, which can be used either to account for static symmetry breaking (\emph{e.g.}\ emergence of long-range polarisation) if tilts are frozen in or to label the corresponding phonon branch, if tilts remain dynamic. The soft-mode description of displacive phase transitions links these two pictures, with the symmetry of the soft phonon dictating the descent in space group symmetry as the tilt distortions become static.\cite{Dove1993}

In addition to conventional inorganic perovskites, there are several molecular perovskite analogues, including organic halide perovskites,\cite{Snaith2013,Mitzi2001} dicyanamides,\cite{Tong2003,Schlueter2004,Schlueter2005,Bermudez-Garcia2016} azides,\cite{Du2015a,Gomez-Aguirre2016} Prussian blue analogues,\cite{Buser1977,Aguila2016} dicyanometallates,\cite{Lefebvre2007,Hill2016} and formates.\cite{Sletten1973,Wang2004a,Wang2004} These are systems of strong scientific currency in which at least one component of the ABX$_3$ perovskite structure is molecular: typically the A-site cation and/or the anionic linker X. An important consequence of the incorporation of molecular components is the emergence of new structural degrees of freedom for which there is no analogue in conventional perovskites. Examples include (i) the so-called ``forbidden'' tilts found in some azides, Prussian Blue analogues, and dicyanometallates, in which neighbouring octahedra (no longer corner-sharing) rotate in the \emph{same} sense as one another,\cite{Du2014,Hill2016,Duyker2016,Kareis2012} and (ii) multipolar order associated with orientational degrees of freedom of molecular A-site cations.\cite{Evans2016,Zhang2015} Coupling of these exotic degrees of freedom to the lattice then allows for entirely new symmetry-breaking mechanisms,\cite{Xu2016a} and hence new crystal engineering strategies for targeting \emph{e.g.}\ multiferroic or NTE responses.\cite{Stroppa2013,Benedek2011,Pitcher2015,Senn2016a}

\begin{figure}
\centering
\includegraphics{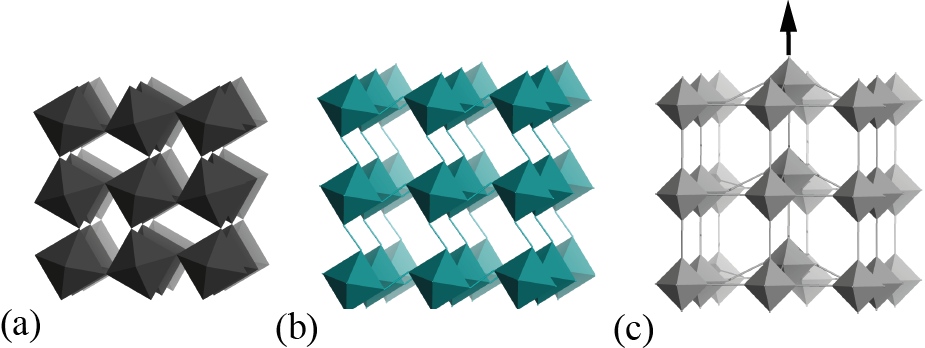}
\caption{Rigid-body distortions in conventional and molecular perovskites. (a) Conventional tilt degrees of freedom, in which neighbouring octahedra rotate in alternating directions. (b) A ``forbidden'' tilt system unique to molecular perovskites in which all coordination octahedra rotate in the same direction. (c) A columnar shift degree of freedom, again unique to molecular frameworks.}
\label{fig1}
\end{figure}

It is natural then to ask: are there any other degrees of freedom of general relevance to the structural chemistry of molecular perovskites? Here, a key consideration is the energy scale associated with different deformations, since those with high energies (\emph{e.g.}\ bond stretches or distortion of coordination geometries) are unlikely to behave as soft modes. Fortunately, simple geometric tools\cite{Giddy1993} can be used to identify distortion modes that preserve bond lengths and coordination geometries and hence are predisposed to play a key role in the low-energy dynamics of materials; these are termed the rigid-unit modes (RUMs) of a given topology. So, for example, it was shown in Ref.~\citenum{Giddy1993} that the only RUMs supported by the conventional perovskite structure are the well-known octahedral tilts discussed above [Fig.~\ref{fig1}(a)]. A similar analysis of the ABX$_3$ lattice with \emph{molecular} X components, however, revealed the existence of \emph{two} types of rotational degrees of freedom (these are the conventional and forbidden tilts) together with a translational degree of freedom involving correlated displacements of columns of connected BX$_6$ octahedra [Fig.~\ref{fig1}(b,c)].\cite{Goodwin2006}

We refer to these columnar translations as `shifts' and argue here that they can indeed play an important role in the structural chemistry of certain families of molecular perovskites. Our paper begins by establishing the conceptual framework for interpreting and characterising columnar shifts. By focusing initially on a two-dimensional simplification of the perovskite framework, we explore the interplay between shift activation and symmetry breaking, the relationship to shear modes, and the potential for coupling with tilt degrees of freedom. We proceed to extrapolate this analysis to the interpretation of static symmetry-breaking distortions in three-dimensional molecular perovskites, drawing on topical case studies from the recent crystal engineering literature. The link to dynamical properties is then made via a simple lattice dynamical model, which we then use to demonstrate that dynamic shift distortions have a distinctive NTE character. Our paper concludes with a discussion regarding the possibility of developing `shift engineering' approaches as an alternative mechanism of accessing polar states in molecular perovskites.



\section{Theory}

\subsection*{Rigid unit modes in molecular perovskites}\label{rums}

Our starting point is a brief recap of the key results in the RUM analysis of Ref.~\citenum{Goodwin2006}; our aim is to clarify the particular periodicities at which shift distortions can occur in molecular perovskites and their 2D analogues. This analysis makes use of a dynamical matrix approach in which a simplified lattice-dynamical model is used to assign energies to distortion patterns.\cite{Giddy1993,Pawley1972} While the absolute energies are not themselves meaningful, the model is set up such that those modes---and only those modes---which do not result in changes to individual bond lengths or coordination environments correspond to solutions with energies exactly equal to zero. Formally, this is achieved using a molecular `rigid unit' translation/rotation basis for the dynamical matrix with the lattice enthalpy determined by the variation in separation between neighbouring rigid units:
\begin{equation}
\phi_{jj^\prime}=\frac{1}{2}K(d_{jj^\prime}-\bar d)^2.
\end{equation}
Here $\phi$ is the interaction potential between neighbouring units $j$ and $j^\prime$, $d_{jj^\prime}$ the corresponding inter-unit separation, $\bar d$ the equilibrium separation, and $K\neq0$ the (fictitious) force constant. Having set up the dynamical matrix $\mathbf D(\mathbf k)$ as in Refs.~\citenum{Giddy1993} and \citenum{Goodwin2006}, the RUMs are identified by the eigenstates of $\mathbf D(\mathbf k)$ for which the corresponding eigenvalue is zero. By varying the distortion periodicity $\mathbf k$, the entire set of RUM-type degrees of freedom can be determined comprehensively.

In practice, the form of $\mathbf D(\mathbf k)$ is really very simple for molecular perovskites. Even so, we consider first the (even simpler) 2D analogue of connected squares, for which
\begin{equation}
\mathbf D(\mathbf k)=\left[\begin{array}{cc|c}1-\cos(2\pi k_x)&0&0\\ 0&1-\cos(2\pi k_y)&0\\\hline 0&0&0\end{array}\right],\label{dmat}
\end{equation}
where $\mathbf k=[k_x,k_y]=k_x\mathbf a^\ast+k_y\mathbf b^\ast$. The rows and columns of $\mathbf D$ index in turn rigid-body translations parallel to $\mathbf a$, rigid-body translations parallel to $\mathbf b$, and rigid-body rotations within the plane; the separation between translational and rotational components in Eq.~\eqref{dmat} is indicated using horizontal and vertical lines. The diagonal form of $\mathbf D$ means that the RUMs can be identified by inspection. No matter what the value of $\mathbf k$, the vector $[0, 0, 1]$ is an eigenvector with zero-valued eigenvalue, and hence rigid-body rotations with arbitrary periodicities are valid RUMs of the system: these distortion modes include both the conventional ($\mathbf k=[\frac{1}{2},\frac{1}{2}]$) and ``forbidden'' ($\mathbf k\neq[\frac{1}{2},\frac{1}{2}]$) tilts described in the introduction [Fig.~\ref{fig1}(x)]. The remaining eigenstates have eigenvalues $1-\cos(2\pi k_x)$  and $1-\cos(2\pi k_y)$ and so correspond to RUMs if and only if $k_x=0$ and/or $k_y=0$. The corresponding eigenvectors $[1, 0, 0]$ and $[0, 1, 0]$ describe rigid-body translations parallel to the $\mathbf a$ and $\mathbf b$ crystal axes, respectively. Taken together, this means that rigid-body translations are allowed so long as individual rows and columns displace along the corresponding row/column axis as a collective object: translations parallel to $\mathbf a$ can correlate with periodicities $\mathbf k=[0,k_y]$ for any $k_y$; those parallel to $\mathbf b$ can correlate with periodicities $\mathbf k=[k_x,0]$.

\begin{figure}
\centering
\includegraphics{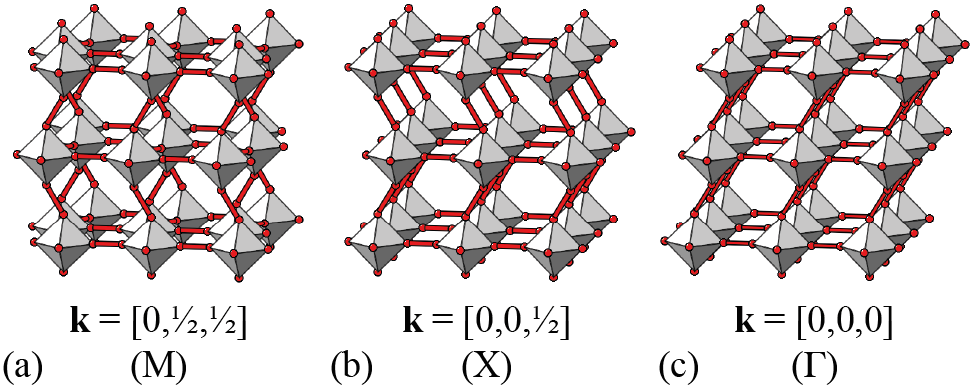}
\caption{Representative shift distortions with different periodicities. In all three cases, shifts occur parallel to the $\mathbf a$ axis (approximately horizontal in this representation). (a) When correlated at $\mathbf k=[0,\frac{1}{2},\frac{1}{2}]$, neighbouring columns shift in alternating directions. (b) At $\mathbf k=[0,0,\frac{1}{2}]$, entire planes of octahedra shift in the sense along $\mathbf a$; the direction of this translation reverses between neighbouring planes. (c) When correlated at $\mathbf k=[0,0,0]$ shift distortions resemble a shear of the framework structure.}
\label{fig2}
\end{figure}

These results translate directly to the three-dimensional case of molecular perovskites. The dynamical matrix now assumes the form
\begin{equation}
\mathbf D(\mathbf k)=\left[\begin{array}{c|c}\begin{array}{ccc}1-\cos(2\pi k_x)&0&0\\ 0&1-\cos(2\pi k_y)&0\\ 0&0&1-\cos(2\pi k_z)\end{array}&\ast\\ \hline \ast&\ast\end{array}\right],
\end{equation}
where the symbol $\ast$ denotes a null $3\times3$ submatrix; the six rows and columns of $\mathbf D$ index first rigid-body translations along the crystal axes $\mathbf a, \mathbf b, \mathbf c$ and then rigid-body rotations about these same axes. The octahedral shift distortions correspond to the first three eigenstates. In each case, the corresponding eigenvalue is zero valued only if the relevant wave-vector component $k_\alpha=0$ (\emph{i.e.}, for shifts parallel to axis $\alpha\in\{\mathbf a, \mathbf b, \mathbf c\}$). Hence, the shift degrees of freedom in 3D molecular perovskites also involve collective row/column displacements polarised along the row/column axis. By way of example, shifts involving translations parallel to $\mathbf a$ can propagate with periodicities $\mathbf k=[0,k_y,k_z]$ for any $k_y,k_z$. In the special case that either $k_y$ or $k_z=0$, these shifts involve collective translations of entire planes of octahedra (the $(001)$ and $(010)$ planes, respectively); in the even more special case $k_y=k_z=0$, the shift mode describes a shear of the lattice, in this instance polarised along $\mathbf a$ [Fig.~\ref{fig2}].

\subsection*{Shifts in 2D: enumeration and symmetry breaking}

\begin{figure*}
\centering
\includegraphics{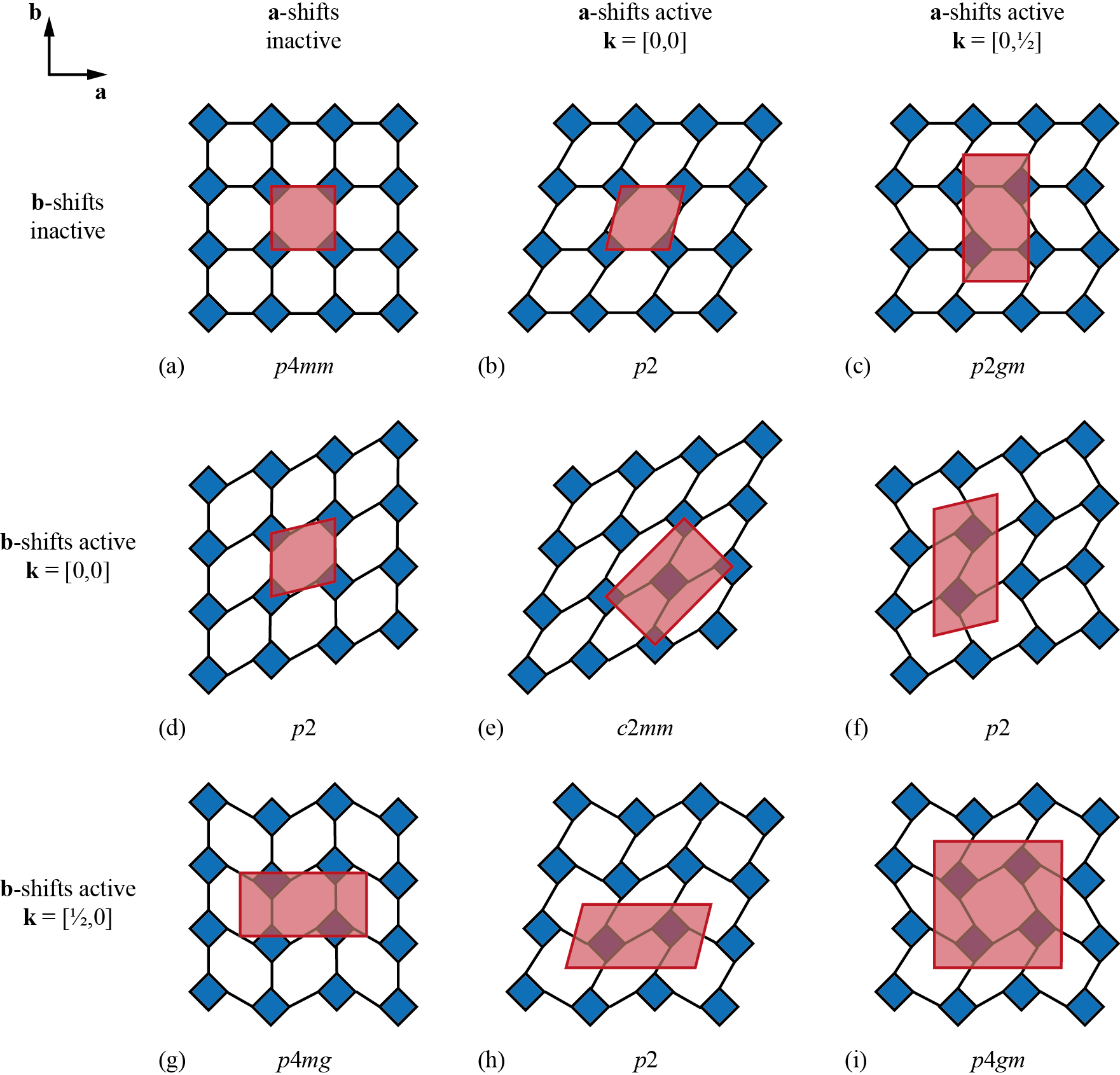}
\caption{Symmetry lowering arising from simple shift distortions of a 2D molecular perovskite analogue. (a) The parent structure has $p4mm$ plane group symmetry. Panels (b)--(i) illustrate the effect of activating zone centre ($\mathbf k=[0,0]$) of zone boundary ($\mathbf k\in\langle\frac{1}{2},0\rangle$) shifts along the $\mathbf a$ and/or $\mathbf b$ axes. The corresponding unit cells are shown in red, and the plane group labels are given below each illustration.}
\label{fig3}
\end{figure*}

We now consider the explicit form of the various possible shift modes for the 2D molecular perovskite analogue of connected squares. The analysis given above indicates that this system supports two types of shifts: one involving collective translations of rows of connected squares along a direction parallel to $\mathbf a$ and modulated with periodicity $\mathbf k=[0,k_y]$; the other involving collective translations of columns of connected squares parallel to $\mathbf b$, where the modulation is now characterised by $\mathbf k=[k_x,0]$. By analogy to the common displacive instabilities in conventional perovskites, we anticipate that the physically most relevant cases are those for which $\mathbf k$ lies either at the zone centre or at the zone boundary---\emph{i.e.}, $k_x,k_y\in\{0,\frac{1}{2}\}$. We limit our analysis to the corresponding set of $\mathbf k$ points, such that for each of the two orthogonal shift systems there are three possibilities: (i) the shifts are inactive, (ii) the shifts are active with $\mathbf k=[0,0]$, or (iii) the shifts are active with $\mathbf k\in\langle0,\frac{1}{2}\rangle$. Since the two sets of shifts are orthogonal this gives us a total of nine cases to consider; we now take these in turn, summarising our discussion in Fig.~\ref{fig3}.

\begin{figure}
\centering
\includegraphics{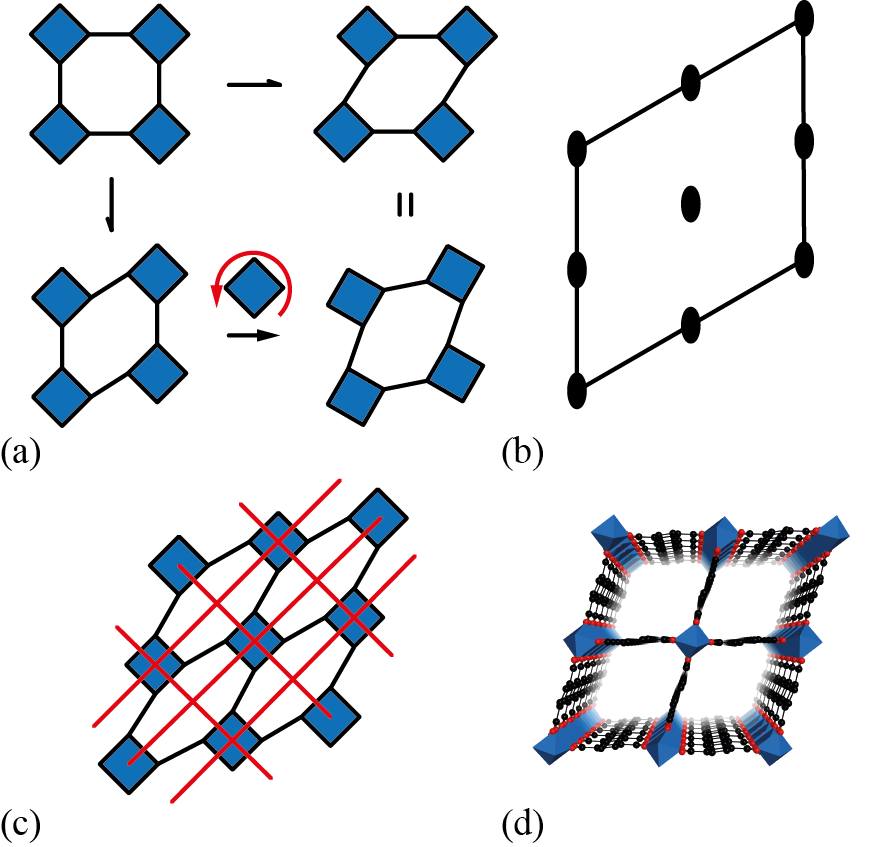}
\caption{Some symmetry relationships in 2D shift systems. (a) Activation of zone-centre shifts leads to structures that are related to one another \emph{via} in-phase tilts of he rigid units. (b) This transformation is continuous because the plane group symmetry elements of the shifted structures (2-fold rotation axes distributed as illustrated here) are compatible with the activation of in-phase tilts. (d) For some shift systems, such as the $c2mm$ distortion shown here, the persistence of mirror symmetry elements forbids mixing of shifts and tilts. (e) This particular shift system is related to the compliant structure of the MIL-53 family, shown here in polyhedral representation.\cite{Nanthamathee2015,Serra-Crespo2015}}
\label{fig4}
\end{figure}

What at face value might appear to be the simplest case---namely, activation of a single shift system with periodicity $\mathbf k=[0,0]$---turns out to give rise to a relatively complex situation. These shifts describe a shear of the perovskite lattice polarised along one of the lattice vectors $\mathbf a$ or $\mathbf b$; the corresponding distortions are illustrated in Fig.~\ref{fig1}(b,d). In both cases the vast majority of the symmetry elements present in the $p4mm$ plane-group symmetry of the parent lattice are lost and the lattice symmetry is now reduced to $p2$. This symmetry lowering is so severe that activation of these shifts allows coupling to an entirely different type of rigid body distortion---namely, the ``forbidden'' (in-phase) tilts also at $\mathbf k=[0,0]$ [Fig.~\ref{fig4}(a,b)]. In fact, these tilts provide a continuous pathway between $\mathbf k=[0,0]$ shifts polarised along $\mathbf a$, on the one hand, and those polarised along $\mathbf b$, on the other hand, such that the former type of shift cannot be distinguished from a combination of the latter shift type together with an in-phase tilt (or \emph{vice versa}). This confusing situation arises because $\mathbf k=[0,0]$ shifts polarised along either $\mathbf a$ or $\mathbf b$ are characterised by the same irreducible representation; in other words, the two shift systems break the parent symmetry in identical ways.

Coupling to tilts is by no means a universal feature of shift distortions, and a counter-example is given by the case in which the two $\mathbf k=[0,0]$ shifts are active to precisely the same extent. This situation corresponds to a shear polarised along the cell diagonal, which results in a much less severe symmetry-lowering process: the resulting plane group is now $c2mm$ [Fig.~\ref{fig3}(e)]. Importantly, the persistence of mirror symmetry elements bisecting the rigid units means that coupling to tilts can only occur by further symmetry lowering [Fig.~\ref{fig4}(c)]. So in this case, the particular shift modes can be uniquely identified from the lattice symmetry. Of course, the transition from $p4mm$ to $c2mm$ structures---couched here in terms of activation of $\mathbf k=[0,0]$ shifts---corresponds to a ferroelastic distortion of the lattice.\cite{Aizu1969,Salje2012a} The ferroelastic state is well known to be mechanically compliant,\cite{Ortiz2012,Coudert2015} and as such is often associated with phenomena such as uniaxial NTE and negative linear compressibility (NLC).\cite{Hunt2015,Cairns2015,Redfern1988} Indeed, the 2D model we consider here may be interpreted as a projection of the 3D ``wine-rack'' structure of well-known compliant framework materials such as the MIL-53 family, which is certainly known to exhibit both NTE and NLC [Fig.~\ref{fig4}(d)].\cite{Nanthamathee2015,Serra-Crespo2015}


Whereas zone-centre shift modes describe ferroelastic distortions, those at the zone boundary give rise to antiferroelastic states. In the case of shifts polarised along $\mathbf a$, the relevant zone boundary periodicity is $\mathbf k=[0,\frac{1}{2}]$. Consequently, activation of this shift mode results in a doubling of the cell in the $\mathbf b$ direction with the corresponding plane group symmetry now $p2gm$. Once again, the persistence of mirror symmetry elements bisecting the rigid units forbids coupling to tilts. The equivalent shift mode polarised along $\mathbf b$ gives rise to an analogous distortion: the cell now doubles along $\mathbf a$ and the plane group symmetry is $p2mg$. In contrast to the situation for the corresponding zone-centre shift modes, in this case there is clearly no continuous pathway between the two states. Simultaneous activation of \emph{both} zone-boundary shift modes to identical extents results in the appealing antiferroelastic distortion shown in Fig.~\ref{fig2}(i). This distortion requires doubling along both cell axes and is described by the plane group $p4gm$. Once again, the point symmetry at the rigid unit site includes a mirror plane and so this particular shift system is symmetry forbidden from coupling with tilt modes. 

For completeness, we consider the final possibility in which a zone-centre shift mode polarised along one axis is combined with a zone-boundary shift polarised along the remaining axis. The corresponding distortions for the two possible axis choices are illustrated in Fig.~\ref{fig2}(f,h). In both cases the cell doubles and in both cases the resulting plane group symmetry is $p2$. Yet, while each shift distortion now has sufficiently low symmetry to couple with tilt modes (as above), there is no continuous path between the two: they are distinguishable by virtue of the particular axis along which the cell has doubled. Our key point in covering all these different possibilities is to demonstrate that activation of different shift modes results in different symmetry-breaking processes that can be fundamentally distinct from those accessible \emph{via} tilt degrees of freedom---whether conventional\cite{Kennedy1999} or forbidden.\cite{Du2014,Hill2016,Duyker2016,Kareis2012}

\subsection*{Notation}\label{notation}

Given the complexity of shift distortions and their combinations---even in 2D---it is clearly desirable to develop a concise notation to represent the particular set of shift modes active in a given structure. In the case of tilt distortions, the most widely-used notation is that of Glazer;\cite{Glazer1972} we first review this notation with the view of subsequently extending the approach to shifts.

Like shifts, independent tilt systems can be associated with each of the three crystal axes. In conventional perovskites, rotations around the $\mathbf a$ axis (by way of example) can propagate with periodicity $\mathbf k=[k_x,\frac{1}{2},\frac{1}{2}]$. Hence, the particular tilt distortion associated with a single axis $\alpha$ is described by two terms: the tilt magnitude $e_\alpha$ and the relevant wave-vector component $k_\alpha$, which---as discussed above---is usually either $0$ (`in-phase' tilts) or $\frac{1}{2}$ (`out-of-phase' tilts). Glazer condenses this information for each axis into a compound symbol $\lambda^\mu$. The index $\mu\in\{0,+,-\}$ denotes whether a tilt is inactive ($\mu=0$; $e_\alpha=0$), in-phase ($\mu=+$; $k_\alpha=0$) or out-of-phase ($\mu=-$; $k_\alpha=\frac{1}{2}$); the primary symbol $\lambda$ reflects the magnitude of an active tilt in order to show the existence or absence of symmetry relationships between tilts along different axes of the parent perovskite lattice. The untilted aristotype has Glazer symbol $a^0a^0a^0$; the term $a^-a^-a^-$ denotes equal-magnitude out-of-phase tilts around each of the three crystal axes; and the term $a^+b^-b^-$ denotes in-phase tilts around $\mathbf a$ with distinct equal-magnitude out-of-phase tilts around $\mathbf b$ and $\mathbf c$. Howard and Stokes established a link between these labels and the corresponding space group symmetries.\cite{Howard1998} We note that the index $\mu$ is equal to the value of $e_\alpha\exp[2\pi{\rm i}k_\alpha]$ if (i) the symbol `$+$' can be associated with 1 and `$-$' with $-1$, and (ii) $e_\alpha$ is taken to equal $1$ for active tilt modes and $0$ for inactive tilt modes.

The various shift distortions of the 2D molecular perovskite structure discussed above are also describable in terms of the magnitude and periodicity of collective translations along each crystal axis. This immediately suggests an analogous notation to that of Glazer's for tilts, with only one subtle conceptual modification: the periodicity implied by the index $\mu$ must now refer to the component of $\mathbf k$ \emph{perpendicular} to the corresponding crystal axis. So, for example, the diagonal ferroelastic distortion discussed in terms of $\mathbf k=[0,0]$ shifts along both $\mathbf a$ and $\mathbf b$ might be summarised by the `Glazer' symbol $a^+a^+$: here the $+$ index would indicate $k_y=0$ for shifts parallel to $\mathbf a$ and $k_x=0$ for shifts parallel to $\mathbf b$; likewise the use of the same primary symbol $a$ would indicate that the shifts have identical magnitude along these two crystal axes. The corresponding symbols for each of the distortions originally presented in Fig.~\ref{fig3} are given in Table~\ref{table1}.

\begin{table}[b]
\small
  \caption{\ A summary of Glazer and matrix notation for the 2D shift systems illustrated in Fig.~\ref{fig3}.}
  \label{table1}
  \begin{tabular*}{0.48\textwidth}{@{\extracolsep{\fill}}llccl}
    \hline
     $\mathbf a$-shifts&$\mathbf b$-shifts&`Glazer'&Matrix&Plane \\
     & & symbol&symbol&group\\
    \hline
    &\\[-5pt]
    inactive&inactive&$a^0a^0$&$\left[\begin{array}{cc}0&0\\ 0&0\end{array}\right]$&$p4mm$\\[10pt]
    in-phase&inactive&$a^+b^0$&$\left[\begin{array}{cc}+&+\\ 0&0\end{array}\right]$&$p2$\\[10pt]
    in-phase&in-phase&$a^+a^+$&$\left[\begin{array}{cc}+&+\\ +&+\end{array}\right]$&$c2mm$\\[10pt]
    out-of-phase&inactive&$a^-b^0$&$\left[\begin{array}{cc}+&-\\ 0&0\end{array}\right]$&$p2gm$\\[10pt]
    out-of-phase&in-phase&$a^-b^+$&$\left[\begin{array}{cc}+&-\\ +&+\end{array}\right]$&$p2$\\[10pt]
    out-of-phase&out-of-phase&$a^-a^-$&$\left[\begin{array}{cc}+&-\\ -&+\end{array}\right]$&$p4gm$\\[10pt]
    \hline
  \end{tabular*}
\end{table}

We will come to show that an unambiguous extrapolation of this notation to 3D molecular perovskites is not straightforward, and so we present an alternative---albeit perhaps more cumbersome---approach similar to that developed in Ref.~\citenum{Hill2016} to describe ``forbidden'' tilts. Here the idea is to exploit the equivalence $\mu\equiv e\exp[2\pi{\rm i}k]$ noted above. We assemble the matrix
\begin{equation}
\left[\begin{array}{ll}\mu_{xx}&\mu_{xy}\\ \mu_{yx}&\mu_{yy}\end{array}\right],\label{mumat}
\end{equation}
where $\mu_{\alpha\beta}\equiv e_\alpha\exp[2\pi{\rm i}k_\beta]$ describes both the magnitude $e_\alpha$ of shift displacements parallel to axis $\alpha$ and also the component $k_\beta$ of the corresponding periodicity $\mathbf k$ parallel to axis $\beta$. We note that if $\beta=\alpha$ then $k_\beta=0$; this is the result of the RUM analysis given above. For consistency we use the Glazer $0,+,-$ symbols for $\mu$ rather than the numerical values of $e_\alpha\exp[2\pi{\rm i}k_\beta]$. So, for the diagonal ferroelastic distortion (assigned Glazer symbol $a^+a^+$ above) we now have the matrix representation
\begin{equation}
\left[\begin{array}{ll}+&+\\ +&+\end{array}\right].
\end{equation}
Equivalent representations for each of the 2D shift distortions are listed in Table~\ref{table1}.

\subsection*{Extension to 3D}

The key result of our RUM analysis was to show that the shift degrees of freedom in 3D molecular perovskites involve collective displacements of columns of octahedra along a direction parallel to the column axis $\alpha$. Shifts may occur along any combination of the three crystal axes; the only constraint on the periodicity $\mathbf k$ of these displacements is that component $k_\alpha$ must equal zero for shifts polarised along axis $\alpha$. Consequently, the shifts associated with each axis now require \emph{three} terms if they are to be described completely: a magnitude $e_\alpha$ together with the \emph{two} wave-vector components $k_\beta,k_\gamma$ corresponding to the \emph{two} axes perpendicular to $\alpha$. It is this additional complexity that renders ambiguous the direct extrapolation of the Glazer notation to 3D shifts.

\begin{figure}
\centering
\includegraphics{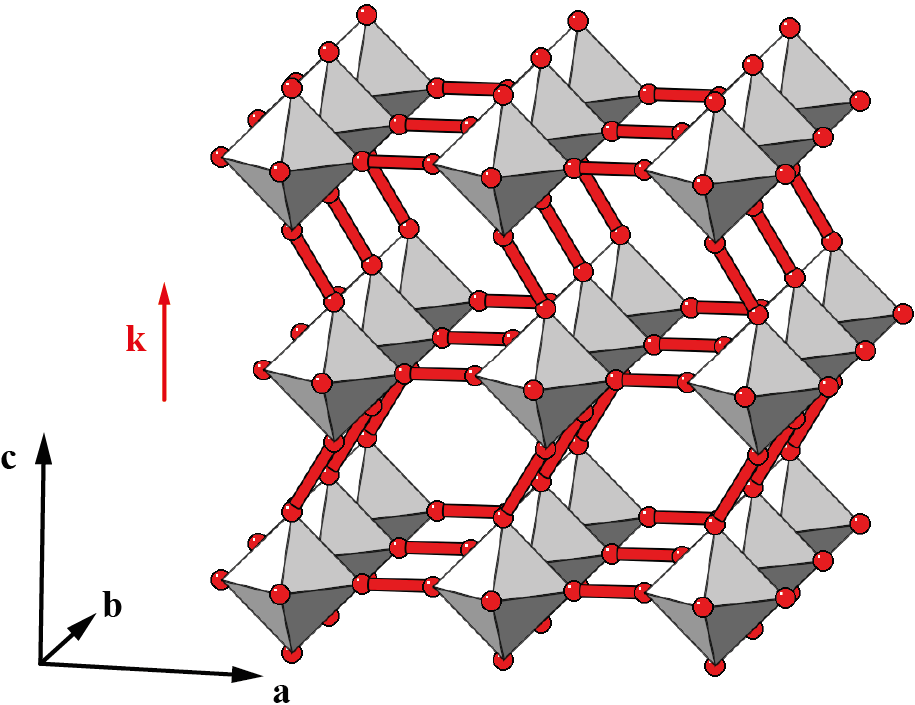}
\caption{An antiferroelastic planar shift system characterised by displacements parallel to $\mathbf a$, correlated with modulation wave-vector $\mathbf k = [0,0,\frac{1}{2}]$.}
\label{fig5}
\end{figure}

By contrast, the more cumbersome matrix notation is straightforwardly extended to 3D shifts: we use the representation
 \begin{equation}
\left[\begin{array}{lll}\mu_{xx}&\mu_{xy}&\mu_{xz}\\ \mu_{yx}&\mu_{yy}&\mu_{yz}\\ \mu_{zx}&\mu_{zy}&\mu_{zz}\end{array}\right],
\end{equation}
defined exactly as for Eq.~\eqref{mumat}. By way of example, the antiferroelastic planar shift distortion shown in Fig.~\ref{fig5} would be characterised by the shift matrix
\begin{equation}
\left[\begin{array}{ccc}+&+&-\\ 0&0&0\\ 0&0&0\end{array}\right].
\end{equation}
Here, the first row signifies that shifts polarised along $\mathbf a$ are active, and propagate with periodicity $\mathbf k=[0,0,\frac{1}{2}]$. The second and third rows signify that shifts along $\mathbf b$ and $\mathbf c$ are inactive. This particular distortion results in symmetry lowering of the $Pm\bar3m$ aristotype to $Pmma$.

One possible approach to modifying the Glazer-type notation for these 3D shifts might be to exploit the Bradley-Cracknell abbreviations for high-symmetry points in the Brillouin zone.\cite{Bradley1972} In some cases, the use of this abbreviation as the Glazer index $\mu$ would allow unambiguous identification of the two required wave-vector components. For example, the $Pmma$ shift system discussed immediately above might be assigned the Glazer symbol $a^{\rm X}b^0c^0$. Here, the index `$\rm X$' of the first term signifies that shifts polarised along $\mathbf a$ are active and are modulated with a periodicity $\mathbf k\in\langle\frac{1}{2},0,0\rangle$. Since $k_x$ must equal zero, we know that $\mathbf k=[0,\frac{1}{2},0]$ or $[0,0,\frac{1}{2}]$; in the absence of active shifts along $\mathbf b$ or $\mathbf c$ these two periodicities give rise to symmetry-equivalent distortions. Despite this success of the nomenclature in this one example, it is straightforward to envisage scenarios in which unambiguous identification is not possible. Nevertheless, for each of the case studies below, we try to give both Glazer and matrix notations, with the understanding that future usage will likely determine limitations of the two approaches and identify of which of these is the more useful in practice.

\section{Case studies}

Having established a theoretical basis with which to identify and categorise shift distortions in molecular perovskites, we proceed to interpret the structures of three experimental systems in this context. Our goals are to demonstrate that a variety of different shift systems is observed experimentally, and to highlight the potential for interplay with tilt and A-site orientational degrees of freedom.

\subsection*{Tetramethylammonium calcium azide}

\begin{figure}[t]
\centering
\includegraphics{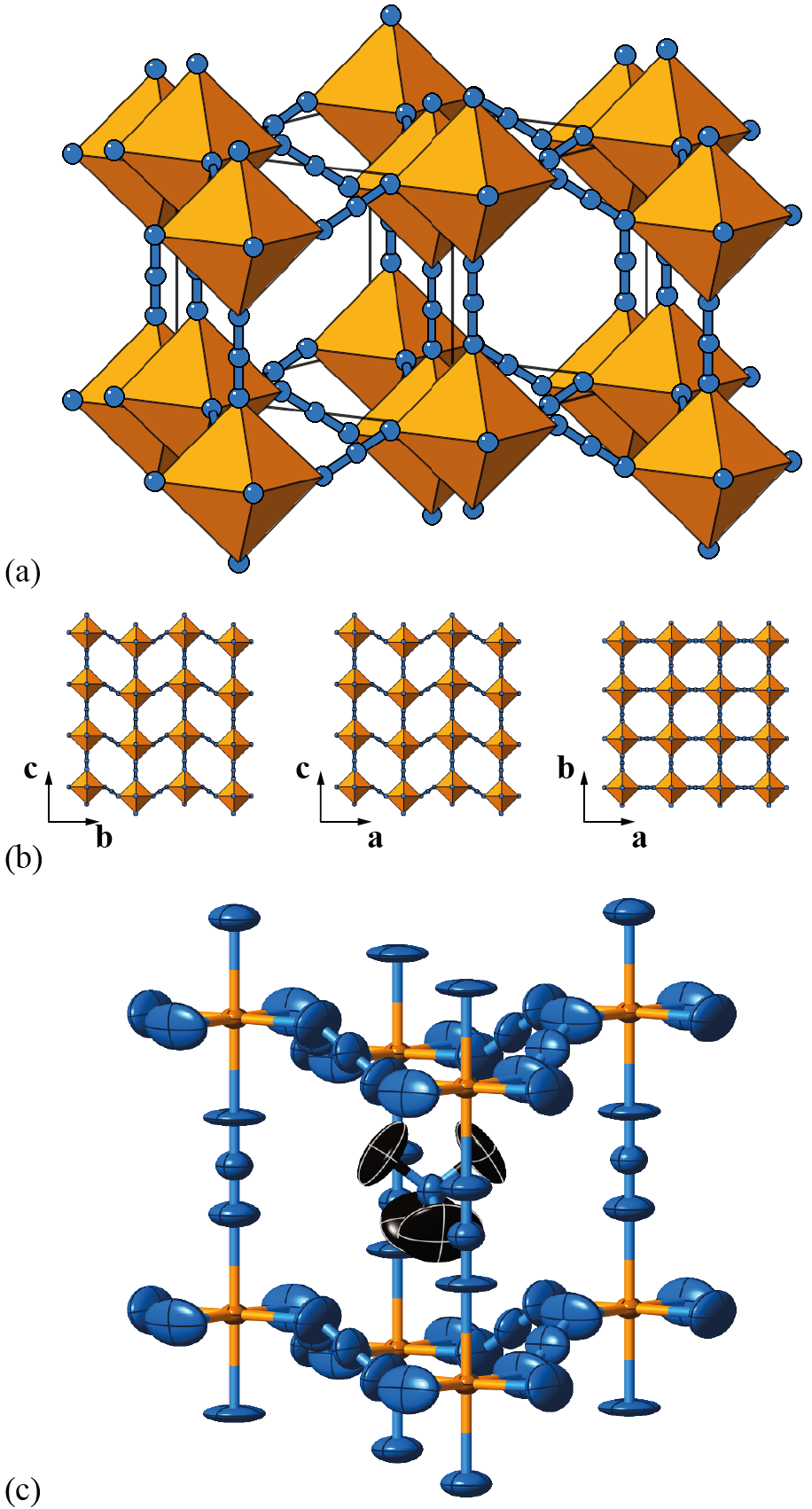}
\caption{Static shift distortions in [NMe$_4$]Ca[N$_3$]$_3$. (a) A polyhedral / ball-and stick representation of the crystal structure of [NMe$_4$]Ca[N$_3$]$_3$, as reported in Ref.~\citenum{Mautner1988}. Ca atoms are shown in yellow and N atoms in blue. The [NMe$_4$]$^+$ cations have been omitted for clarity. Shifts are polarised along $\mathbf c$ (the vertical axis in this representation) and are related to those illustrated originally in Fig.~\ref{fig2}(a). (b) 2D sections of the crystal structure using the same representations as in (a). These sections lie perpendicular to the $\mathbf a$, $\mathbf b$, and $\mathbf c$ axes (left--right) and relate the 2D shift systems enumerated in Fig.~\ref{fig3} with the matrix representation of the full 3D shift system active in this material. (c) A representation of the local environment of the [NMe$_4$]$^+$ cation in this material; colours are as for (a) and (b), with C atom shown in black. Thermal ellipsoids are given at 50\% probability. There is a close match in A-site cation geometry and the framework distortion effected by shift activation. The relatively large thermal ellipsoids suggest substantial dynamic disorder in this system. \label{fig6}}
\end{figure}

Our first example is the azide-containing perovskite framework [NMe$_4$]Ca[N$_3$]$_3$ (Me = CH$_3$), the structure of which was reported in Ref.~\citenum{Mautner1988}. At room temperature, this system adopts a tetragonal structure (space group $P4/nmm$) with cell parameters related to that of the cubic aristotype by $a\sim\sqrt2a_0$ and $c\sim a_0$ [Fig.~\ref{fig6}(a)]. It can be shown that this symmetry is entirely accounted for by the presence of an active shift system along the tetragonal axis.\cite{Campbell2006} In this particular example, columns of CaN$_6$ octahedra aligned parallel to $\mathbf c$ are shifted along $\mathbf c$ relative to their immediate neighbours. The shift pattern alternates along $\mathbf a$ and $\mathbf b$ such that the distortion is clearly associated with the modulation wave-vector $\mathbf k=[\frac{1}{2}, \frac{1}{2}, 0]$ (given relative to the parent cell). There are no shifts along either $\mathbf a$ or $\mathbf b$. So, using the approaches described above, we identify this distortion with the Glazer symbol $a^0a^0c^{\rm M}$ and the matrix representation
\begin{equation}
\left[\begin{array}{ccc} 0&0&0\\ 0&0&0\\ -&-&+\end{array}\right].
\end{equation}
As an aside, we note that one straightforward method of assigning tilts is to consider cross-sections taken perpendicular to each parent axis, from which the corresponding 2D shifts may be determined by inspection [Fig.~\ref{fig6}(b)]:
\begin{equation}
\left[\begin{array}{ccc} \cdot&\cdot&\cdot\\ \cdot&0&0\\ \cdot&-&+\end{array}\right]\qquad\left[\begin{array}{ccc} 0&\cdot&0\\ \cdot&\cdot&\cdot\\ -&\cdot&+\end{array}\right]\qquad\left[\begin{array}{ccc} 0&0&\cdot\\ 0&0&\cdot\\ \cdot&\cdot&\cdot\end{array}\right].
\end{equation}
The corresponding 3D shift matrix is a superposition of these three 2D sub-matrices, with the understanding that shifts may sometimes appear inactive in one cross-section but are obviously active in another.

As in a number of the simple 2D cases studied above, the particular shift distortion mode we observe in [NMe$_4$]Ca[N$_3$]$_3$ retains a number of the mirror symmetry elements of the aristotype, which has the effect of preventing mixing between shifts and octahedral tilts. Indeed, there are no static active tilts in the reported structure. What is clear, however, is that there is likely a large degree of dynamic distortion, given the magnitude of the thermal ellipsoids. Consequently, it is possible that this system will exhibit displacive phase transitions on cooling; a re-examination using variable temperature methods may be rewarding in this case.

But what drives the presence of static shifts in this system? We offer two suggestions. The first concerns the coordination preference of the azide anion as a bridging linker. It has long been known that the preferred `end-to-end' bridging geometry involves substantially bent M--N--N angles; together with the \emph{trans}-EE coordination of the N$_3^-$ ion this is presumably what allows such large ($\simeq1.3$\,\AA) displacements between neighbouring Ca$^{2+}$ ions [Fig.~\ref{fig5}(x)]. Indeed this propensity of azide to allow activation of shifts is likely a general phenomenon; however, this point does not explain why it is this particular $a^0a^0c^{\rm M}$ shift system that is adopted here. So our second observation concerns the relationship between the geometry of the [NMe$_4$]$^+$ cation and the structural distortions to the A-site cavity that occur as a result of columnar shifts. In the aristotype structure, the point symmetry at the A site is $m\bar3m$ ($O_h$), which is a supergroup of the $\bar43m$ ($T_d$) symmetry of tetramethylammonium; consequently the cation must exhibit orientational disorder in this parent structure. On activation of the $a^0a^0c^{\rm M}$ shifts, the A-site point symmetry is reduced to $\bar42m$ ($D_{2d}$), a subgroup of $\bar43m$. This allows orientational order of the cation. Indeed, there is a close match between the geometry of the (ordered) cation and the shape of the A-site cavity that suggests the distortions is driven largely by packing and cation--framework interactions [Fig.~\ref{fig6}(c)].

\subsection*{Dimethylammonium manganese azide}

\begin{figure}[b]
\centering
\includegraphics{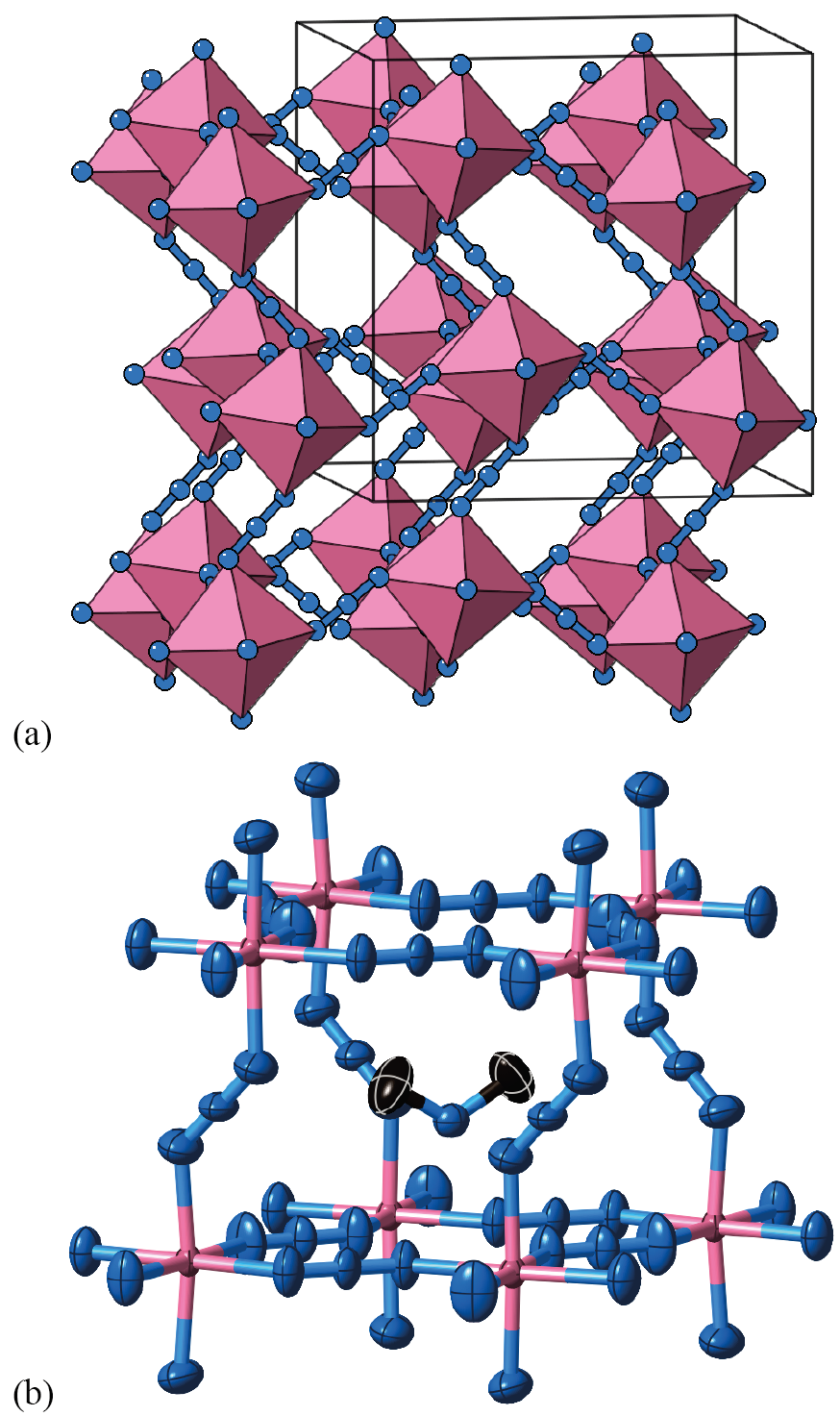}
\caption{Static shift distortions in [NMe$_2$H$_2$]Mn[N$_3$]$_3$. (a) A representation of the crystal structure as reported in Ref.~\citenum{Zhao2013}. Mn atoms are shown as pink polyhedra; N atoms as blue spheres. There are two orthogonal shift distortions active in this system. One is precisely the same as thoat shown in Fig.~\ref{fig6}(a) and gives rise to alternating columnar displacements polarised along $\mathbf c$ (the vertical axis in our representation here). At the same time there is an antiferroelastic planar shift system polarised along the $\mathbf b$ direction (the horizontal axis in this representation) that by itself gives rise to the type of distortion shown in Fig.~\ref{fig5} (albeit with axes relabelled). (b) Thermal ellipsoid representation of the local environment of [NMe$_2$H$_2$]$^+$ cations within the distorted perovskite framework. \label{fig7}}
\end{figure}

A closely related system that supports two orthogonal shift systems at once is the ambient phase of [NMe$_2$H$_2$]Mn[N$_3$]$_3$.\cite{Zhao2013} The reported crystal structure has orthorhombic $Cmce$ symmetry with $a\sim2a_0$, $b\sim2a_0$, $c\sim2a_0$ [Fig.~\ref{fig7}(a)]. The two shift systems present involve displacements along $\mathbf c$ and $\mathbf b$. The former is of precisely the same form as in [NMe$_4$]Ca[N$_3$]$_3$; the latter involves concerted alternating displacements of sheets of octahedra and is associated with the modulation wave-vector $\mathbf k=[0,0,\frac{1}{2}]$. Once again, it can be shown that these two distortions acting together account entirely for the observed space group symmetry;\cite{Campbell2006} in other words, their combined effect acts as the primary order parameter. Using the approach of section \ref{notation} we assign to this distortion the Glazer label $a^0b^{\rm X}c^{\rm M}$ and the shift matrix
\begin{equation}
\left[\begin{array}{ccc} 0&0&0\\ +&+&-\\ -&-&+\end{array}\right].
\end{equation}
This is an example of the ambiguity of the Glazer-type notation we have proposed. In our label $a^0b^{\rm X}c^{\rm M}$ it is not clear whether the shifts polarised along $\mathbf b$ are associated with periodicity $\mathbf k=[0,0,\frac{1}{2}]$ or $[\frac{1}{2},0,0]$; yet these two cases now result in meaningfully different symmetry-breaking processes. In contrast, the matrix representation is unambiguous. What should be immediately apparent from both notations, however, is the existence of a group--subgroup relationship between the structure type of this compound and that of the preceding example.

The arguments presented to explain the activation of $a^0a^0c^{\rm M}$ shifts in [NMe$_4$]Ca[N$_3$]$_3$ appear to hold again for the $a^0b^{\rm X}c^{\rm M}$ shifts we find in [NMe$_2$H$_2$]Mn[N$_3$]$_3$. Clearly the azide linker is common to both, but we find also that the point symmetry at the A site is reduced in order allow orientational order of the [NMe$_2$H$_2$]$^+$ cation [Fig.~\ref{fig7}(b)]. The crystallographic point symmetry of this site is $2$ ($C_2$) in the $Cmce$ structure, which is clearly a subgroup of the idealised $222$ ($C_{2v}$) molecular point symmetry.

One effect of the activation of multiple shift systems is that the crystal symmetry is now sufficiently low that a set of octahedral tilts couples to the shift-induced distortions. This tilt system is characterised by the (conventional) Glazer label $a^-b^0b^0$ and cannot by itself account for the $Cmce$ symmetry. In other words, octahedral tilts do not act as the primary order parameter in this system.

\subsection*{Dimethylammonium manganese formate}

\begin{figure}
\centering
\includegraphics{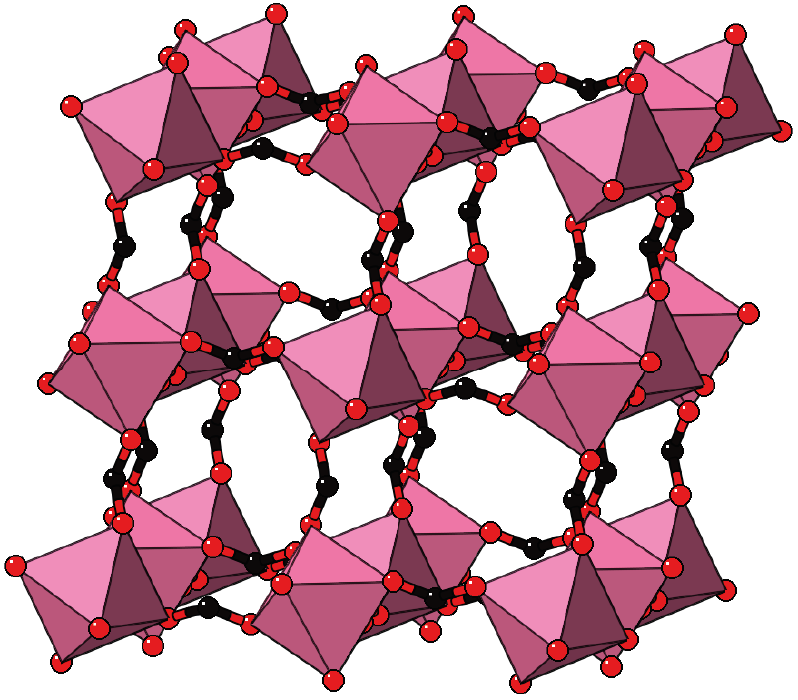}
\caption{Representation of the crystal structure of [NMe$_2$H$_2$]Mn(HCOO)$_3$, as reported in Ref.~\citenum{Wang2004a}. Mn atoms are shown as pink polyhedra, O atoms as red spheres, and C atoms as black spheres. Here the shift distortions are associated with macroscopic shear of the lattice.\label{fig8}}
\end{figure}

In our final case study, we consider a system for which shift distortions are present but clearly not the primary order parameter: [NMe$_2$H$_2$]Mn(HCOO)$_3$. In the high-temperature phase of this compound, the crystal symmetry is $R\bar3c$ with $a\sim\sqrt2a_0$ and $c\sim2\sqrt3a_0$ [Fig.~\ref{fig8}].\cite{Wang2004a} The existence of a rhombohedral distortion itself implies activation of a shear strain polarised along the body diagonal of the ABX$_3$ cube. Consequently the shift distortion is given by the straightforward labels $a^\Gamma a^\Gamma a^\Gamma$ and
\begin{equation}
\left[\begin{array}{ccc} +&+&+\\ +&+&+\\ +&+&+\end{array}\right].
\end{equation}
This distortion reduces the $Pm\bar3m$ aristotype symmetry to $R\bar3m$ ($a\sim\sqrt2a_0$ and $c\sim\sqrt3a_0$), which is a minimal supergroup of the observed space group $R\bar3c$ ($c\sim2\sqrt3a_0$) and so cannot act as a primary order parameter. Instead, it is the conventional octahedral tilt distortion (Glazer notation $a^-a^-a^-$) that is responsible for breaking the aristotypic symmetry; here the shifts couple to the tilts.

\subsection*{Some complications}

In selecting these case studies we have intentionally focussed on systems for which the active shift distortions are relatively straightforward. There is absolutely no difficulty in anticipating complicating factors in other systems that would make the kind of analysis we present much trickier. We briefly highlight some of these factors here, noting that many of these are complications also in the characterisation of octahedral tilts in conventional perovskites.

First, there will be systems for which the difference in magnitude of shifts for different directions will meaningfully affect the symmetry of the distorted state. Glazer notation allows this distinction to be made through the use of different primary symbols $\lambda$; however, the matrix notation as presented would need to be modified to reflect this variation---perhaps through the use of variables or constants $\neq\pm1$ in the matrix itself. Second, we have focused on shifts characterised by periodicities at the zone centre or zone boundary. More complex periodicities are allowed: an example occurs in the material [NPr$_4$]Ni(dca)$_3$ (Pr = C$_3$H$_7$; dca = [N(CN)$_2$]$^-$), for which $\mathbf c$-shifts are active and modulated by the wave-vector $\mathbf k=[\frac{1}{4}, \frac{1}{4}, 0]$.\cite{Schlueter2005} One might anticipate the use of the Bradley-Cracknell symbol $\Sigma$ in the corresponding Glazer notation; likewise there is in principle no reason why complex (or in this case, imaginary) values of $\exp[2\pi{\rm i}k_\alpha]$ might not be used in the matrix notation. Nevertheless, in both cases there are issues of distortion \emph{phase} that are probably too difficult to be unambiguously resolved by a terse symbolic representation. And, third, it is perfectly feasible for a system to support more than one shift distortion along a given axis. Indeed, this may not be particularly rare, given that zone-centre shifts correspond to shear modes.\cite{Zhao2013} This situation is akin to the well-known case of `compound tilts' found in the study of some inorganic perovskites.\cite{Peel2012} 

\section{Dynamic shifts}

So far, our focus has been on the characterisation and understanding of static shift distortions in molecular perovskites. For a structural degree of freedom---such as shifts---to influence the phase behaviour of broad family, its effect on the lattice dynamics is an equally important consideration. The field would surely gain from experimental studies of the lattice dynamics in molecular framework analogues (noting, for example, the transformative role played by inelastic neutron scattering in developing the soft-mode theory of phase transitions in conventional perovskites).\cite{Cowley1964,Shirane1974,Dove2002} Our approach here, however, is to develop an extremely simple computational lattice-dynamical model, from which we calculate the corresponding phonon dispersion curves. Through interrogation of the corresponding eigenvectors, we are able to explore the role of shift modes in the phonon spectrum for this representative model. We proceed to calculate the corresponding Gr{\"u}neisen parameters---a measure of the role of individual modes in the thermal expansion behaviour of a material\cite{Gruneisen1912}---and demonstrate that dynamic shifts may play an as-yet under-appreciated role in the NTE properties of some molecular perovskites.

\subsection*{Lattice-dynamical model}

\begin{figure}
\centering
\includegraphics{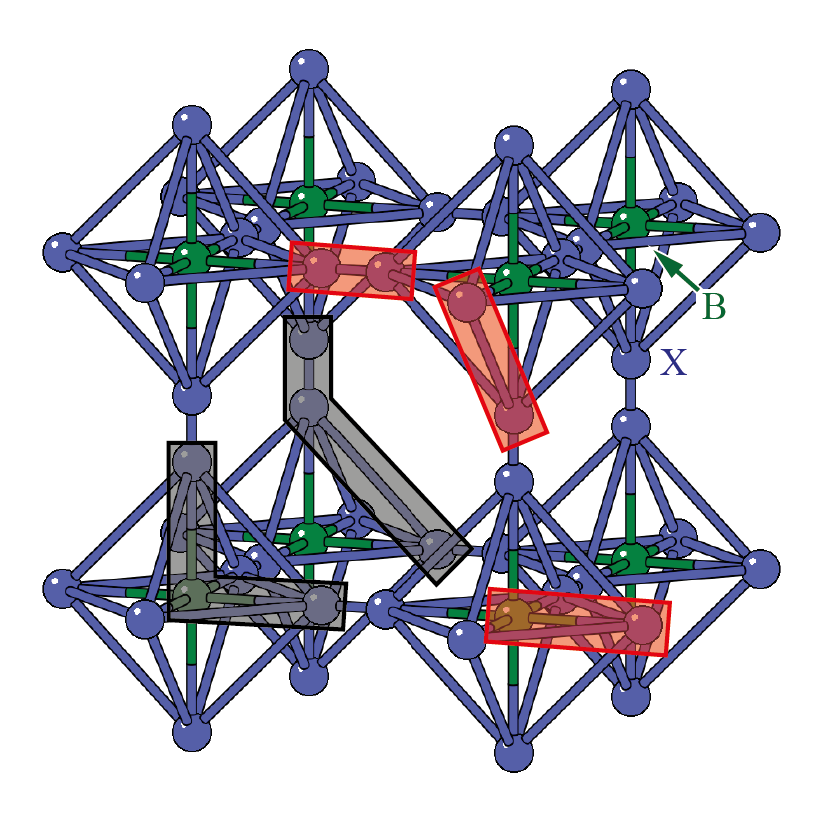}
\caption{Representation of the simplified molecular perovskite structure used in our lattice dynamical model. B atoms (shown in green) are located at the origin of the $Pm\bar3m$ cell, and X atoms (shown in blue) are located on the $6e$ site. The network is held together by a combination of harmonic bond stretching potentials (relevant pairs highlighted in red) and bond angle potentials (relevant triplets highlighted in black).\label{fig9}}
\end{figure}

The essential features of the simple lattice-dynamical model we develop to study shift distortions are: (i) a perovskite topology, (ii) molecular linkers within the perovskite framework, (iii) rigid metal--linker and intra-linker bonds, (iv) rigid linker--metal--linker bond angles, and (v) flexible metal--linker--metal bond angles. We satisfy these criteria with a simple cubic cell (symmetry $Pm\bar3m$) containing a single atom (`B') at the cell origin, and a linker atom (`X') at the $6e$ site $(x,0,0)$ with $x=0.4$. This model has nominal composition B(X$_2$)$_3$. Because we are primarily concerned with deformations of the framework lattice, and because we want to keep our model as simple as possible, we do not include an A-site cation, and we treat both B and X atoms as charge neutral. Our model is made elastically stable through the introduction of harmonic bond-length and bond-angle interactions, as represented in Fig.~\ref{fig9}. Consequently, the lattice enthalpy of our model is given by
\begin{equation}
E_{\textrm{latt}}=\frac{1}{2}\sum_{\textrm{bonds}}k_{\textrm{harm}}(r-r_0)^2+\frac{1}{2}\sum_{\textrm{angles}}k_{\textrm{angle}}(\theta-\theta_0)^2.
\end{equation}
The relevant parameters for a stable implementation of this model within the GULP program (Ref.~\citenum{Gale1997}) are given in Table~\ref{table2}. Calculations were carried out at fixed volume and checked thoroughly for convergence. Our use of a model for which the equilibrium B--X--X angle is 180$^\circ$ is entirely intentional: this is our mechanism of ensuring shift degrees of freedom are reflected in the phonons rather than in static distortions.

\begin{table}[b]
\small
  \caption{\ Parameter values for the lattice dynamical model described in the text and implemented in GULP.\cite{Gale1997}}
  \label{table2}
  \begin{tabular*}{0.48\textwidth}{@{\extracolsep{\fill}}lll}
    \hline
    Parameter&value\\
    \hline
    Space group&$Pm\bar3m$\\
    $a$ (\AA)&5.0\\
    $m$(B) (a.m.u.)&54.94\\
    $m$(X) (a.m.u.)&16.00\\
    $k_{\textrm{harm}}$(B--X) (eV/\AA$^2$)&1.0\\
    $r_0$(B--X) (\AA)&2.0\\
    $k_{\textrm{harm}}$(X--X) (eV/\AA$^2$)&1.0\\
    $r_0$(X--X) (\AA)&1.0\\
    $k_{\textrm{harm}}$(X$\ldots$X) (eV/\AA$^2$)&1.0\\
    $r_0$(X$\ldots$X) (\AA)&2.828\\
    $k_{\textrm{angle}}$(X--B--X) (eV/rad$^2$)&1.0\\
    $\theta_0$(X--B--X) ($^\circ$)&90.0\\
    $k_{\textrm{angle}}$(X--X--X) (eV/rad$^2$)&0.01\\
    $\theta_0$(X--B--X) ($^\circ$)&135.0\\
    \hline
  \end{tabular*}
\end{table}

\begin{figure}
\centering
\includegraphics{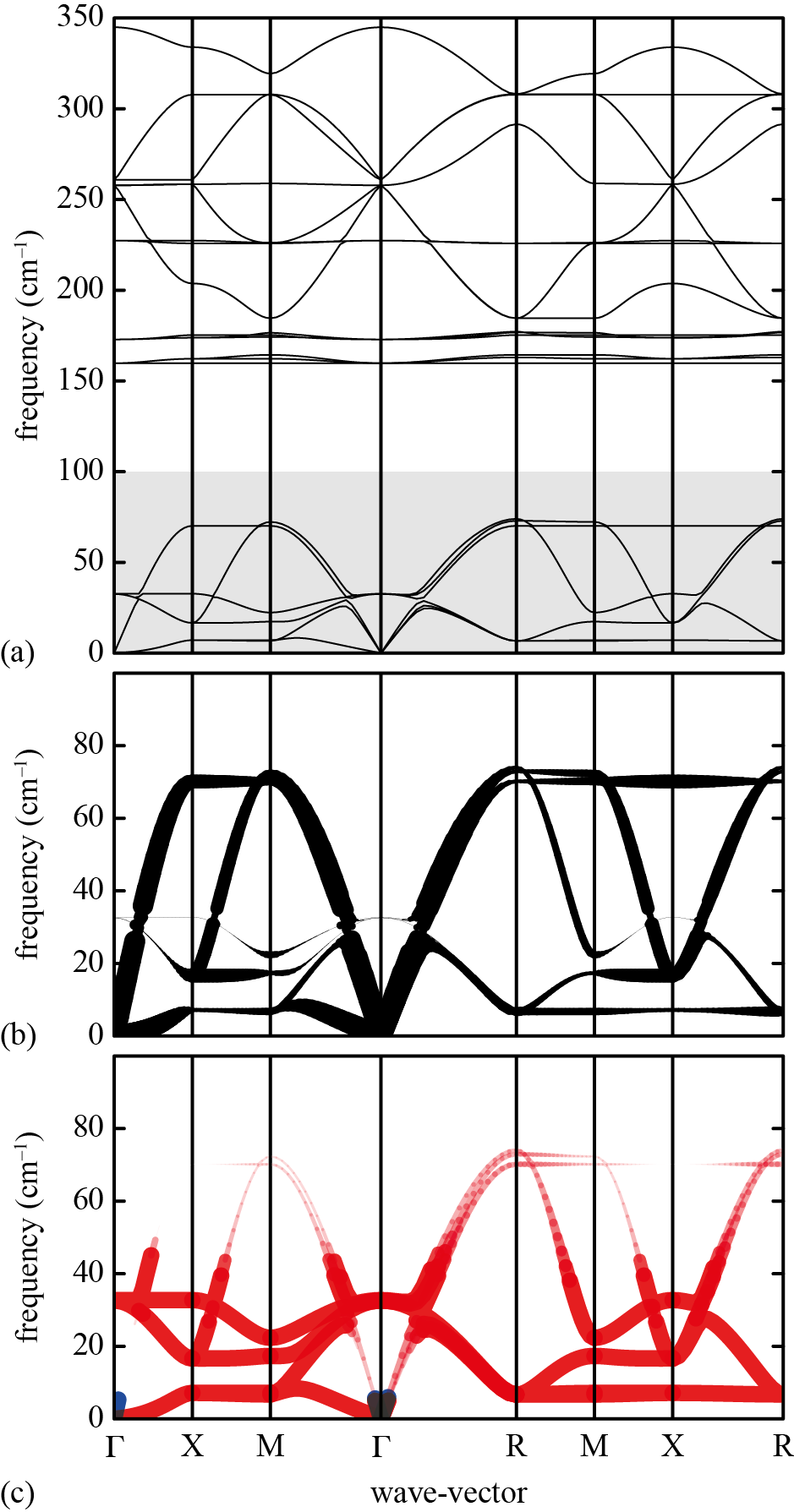}
\caption{Phonon dispersion curves for our lattice dynamical model and their interpretation in terms of shift degrees of freedom. (a) The entire phonon dispersion across selected high-symmetry directions in reciprocal space, as determined using GULP.\cite{Gale1997} The shaded region at frequencies below 100\,cm$^{-1}$ contains the modes responsible for NTE behaviour. (b) The low-frequency region of the phonon dispersion (as shown in (a)) where the branches have been broadened according to the corresponding value of $\rho(\mathbf k,\nu)$. Consequently, those branches that appear bold in this representation correspond to modes with significant translational components.  (c) The same low-frequency region of the phonon spectrum now coloured and broadened according to the value of the mode Gr{\"u}neisen parameter: blue values correspond to $\gamma>0$ and red to $\gamma<0$. The branches that appear bold and red are the most important for NTE; our key result is that these include the shift modes as discussed in the text. \label{fig10}}
\end{figure}

We proceeded to calculate the harmonic phonon dispersion relation for this simple lattice-dynamical model, making use of a $k$-grid of roughly 0.025 reciprocal lattice units. The corresponding phonon dispersion curves along specific high-symmetry directions are shown in Fig.~\ref{fig10}(a). We do not attach any significance to the absolute energy scale of these excitations, since we have not aimed to replicate experimental values in our choice of harmonic spring constants. What is significant is the partitioning of the spectrum into a low-energy regime (which we will come to show dominates NTE behaviour) and a higher-energy regime. With respect to the low-energy component, we note the anomalous slope of the transverse acoustic branch along the $\Gamma$--X direction that is diagnostic of a shear instability, the existence of multiple dispersionless bands (evidence of localised degrees of freedom), and also the presence of zone-boundary soft modes.

In order to better understand the distribution of shift modes throughout this phonon dispersion, we exploited the observation that shifts are associated with eigenvectors $\mathbf e(\mathbf k,\nu)$ uniformly polarised along a single Cartesian axis. Consequently, the projections
\begin{equation}
\rho(\mathbf k,\nu)=\sum_{\alpha}\left|\sum_j\frac{1}{\sqrt{m_j}}e_{j\alpha}(\mathbf k,\nu)\right|^2
\end{equation}
are proportional to the extent to which each mode $\nu$ at wave-vector $\mathbf k$ corresponds to collective translations. Here $\alpha$ indexes the Cartesian axes and $j$ indexes the atoms in the unit cell. In Fig.~\ref{fig10}(b) we show the same low-energy phonon dispersion curves as in Fig.~\ref{fig10}(a) but where we have broadened the curves according to the corresponding value of $\rho$. This highlights visually the distribution of shifts throughout the low-energy phonon spectrum.

What is immediately clear is that shifts play an active role in the low-energy dynamics for those branches along which they are allowed. For the $\Gamma$--X direction, by way of example, the soft acoustic branch is almost entirely accounted for in terms of shift distortions. This branch is doubly degenerate; its low energy reflects the ease with which planar shifts can be accommodated in this simple model. As $\mathbf k\rightarrow$ X, this branch anti-crosses with a rotational RUM branch, such that at the X point itself the shifts correspond to the set of modes with the second lowest phonon frequencies. Note that the longitudinal acoustic branch has increased significantly in energy at this point, such that translations polarised along the same direction as $\mathbf k$ have very much higher energies. Across the X--M direction, one of the two shift degrees of freedom accessible at X becomes increasingly stiff, such that at M itself there is only one shift degree of freedom remaining at the lowest energies. This degree of freedom couples strongly with the rotational RUMs such that it contributes to all three lowest-energy phonon branches. These observations are entirely consistent with the RUM analysis of section~\ref{rums}.

\subsection*{Negative thermal expansion}

The phonon spectrum is directly linked to NTE behaviour via the Gr{\"u}neisen parameters
\begin{equation}
\gamma(\mathbf k,\nu)=-\frac{\partial\ln\omega}{\partial\ln V},
\end{equation} 
where $\omega$ is the frequency of mode $\nu$ at wave-vector $\mathbf k$, and $V$ is the unit cell volume. NTE is driven by those modes for which $\gamma$ is large and negative, especially if these also occur at the very lowest energies.\cite{Evans1999a} We determined the variation in $\gamma$ across our phonon spectrum by applying a 1\% strain to our lattice-dynamical model and recalculating the corresponding $\omega(\mathbf k,\nu)$ values. In Fig.~\ref{fig10}(c) we show the low-frequency region of the phonon dispersion where we have coloured (and broadened) the dispersion curves according to the magnitude and sign of $\gamma$. We find the vast majority of low-energy modes are capable of driving NTE, including the branches associated with shift degrees of freedom. In fact it is possible to count the number of key NTE modes at each of the high-symmetry points (taking care to account for branch degeneracy as appropriate): there are six for $\mathbf k\rightarrow\Gamma$, five at X, four at M, and three at R. In each case, three of these modes correspond to rotational degrees of freedom. So the tilt modes usually used to explain NTE behaviour in perovskite analogues (\emph{e.g.}\ the Prussian Blues\cite{Goodwin2005a,Chapman2006,Adak2011}) are certainly relevant. But our analysis here shows definitively that they need not be the only modes contributing strongly to NTE, and instead correlated shifts can also play a key role. This result reflects our current understanding of NTE in the canonical metal--organic framework MOF-5,\cite{Rimmer2014} the structure of which might reasonably be considered analogous to an A-site deficient molecular perovskite.

The combination of large negative Gr{\"u}neisen parameters and low phonon frequencies also suggests shift-type vibrational modes are likely to show strongly anharmonic behaviour. Hence, the soft mode instabilities normally associated with octahedral tilts and/or ferroelectric displacements may also involve correlated shifts in perovskite analogues. The equilibrium geometry of the B--X--X angle and the presence and charge distribution of A-site cations will help shape the phonon dispersion and---by virtue of the close match in A-site geometry and perovskite deformation noted in the various case studies above---might also be expected to drive phonon softening in suitable cases. Molecular dynamics studies, such as those used to interrogate negative thermal expansion in Zn(CN)$_2$,\cite{Trousselet2015,Fang2013} would provide valuable insight into the possible existence and phenomenology of displacive transitions involving shift degrees of freedom.



\section{Concluding remarks}

So our study has demonstrated that shift distortions are \emph{bona fide} structural degrees of freedom in molecular perovskite analogues: they influence the crystal structures of a number of known compounds, and they are likely to play a key role in the low-energy dynamics of these systems, including anomalous thermal responses such as NTE. We have shown the potential for interplay between shift and tilt degrees of freedom, and again between shift and A-site orientational (multipolar) degrees of freedom; consequently, variation in cation size and geometry might be expected to allow control over the selective activation of specific shift distortions.

\begin{figure}
\centering
\includegraphics[width=8.3cm]{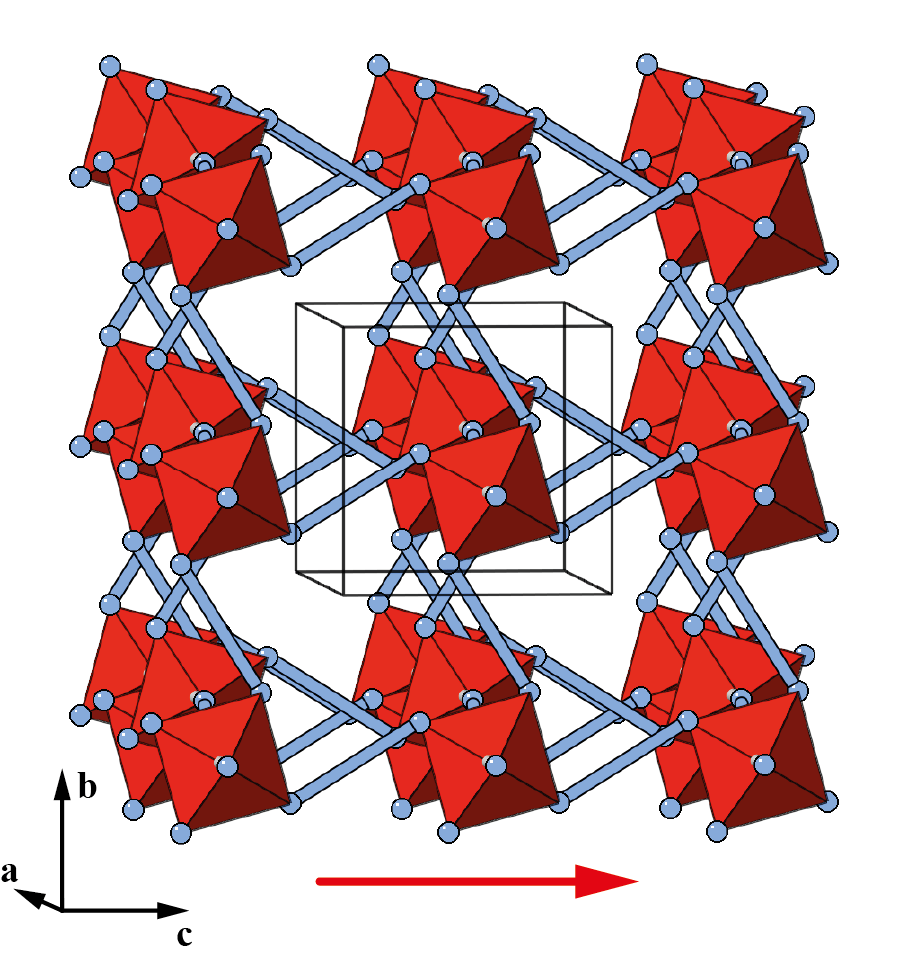}
\caption{Representation of a hypothetical polar $Pma2$ molecular perovskite phase, where the polarisation is indirectly induced via superposition of shift and tilt distortions. Here the blue rods signify molecular bridging units of indeterminate composition. The polarisation direction is indicated by the red arrow.\label{fig11}}
\end{figure}

A crucial result of our study  has been to show that shift distortions can give rise to symmetry-lowering processes inaccessible through \emph{e.g.}\ octahedral tilt mechanisms. The importance of this result lies in the emerging interest in exploiting compound distortions as indirect mechanisms of driving polarisation:\cite{Oh2015,Stroppa2011,Stroppa2013} these are the strategies of so-called ``tilt engineering'', which is allowing access to entirely new families of multiferroic materials.\cite{Pitcher2015} The new symmetry-breaking mechanisms we identify here allow in principle for analogous ``shift engineering'' approaches, where combinations of various correlated shifts---perhaps coupled with tilt or cation order---might be used to break inversion symmetry. For instance, we find that the combination of $a^{\rm X}b^0c^0$ shifts characterised by the matrix
\begin{equation}
\left[\begin{array}{ccc} +&+&-\\ 0&0&0\\ 0&0&0\end{array}\right]
\end{equation}
acting together with $[001]$ tilts propagating at $\mathbf k=[0, 0, \frac{1}{2}]$ gives a distorted structure with polar space group symmetry $Pma2$ [Fig.~\ref{fig11}]. This mechanism of breaking inversion symmetry is strongly reminiscent of the effect of A-site cation order in some AA$'$B$_2$O$_6$ double perovskites;\cite{Rondinelli2012} the key difference of course is that shift or tilt distortions might readily be inverted under the influence of an alternating electric field, whereas cation order is much more difficult to influence.

\section*{Acknowledgements}

J.A.H. and A.L.G. gratefully acknowledge funding through the European Research Council (Grant 279705) and the E.P.S.R.C.

\balance

\bibliography{pccp_2016_shifts} 

\providecommand*{\mcitethebibliography}{\thebibliography}
\csname @ifundefined\endcsname{endmcitethebibliography}
{\let\endmcitethebibliography\endthebibliography}{}
\begin{mcitethebibliography}{63}
\providecommand*{\natexlab}[1]{#1}
\providecommand*{\mciteSetBstSublistMode}[1]{}
\providecommand*{\mciteSetBstMaxWidthForm}[2]{}
\providecommand*{\mciteBstWouldAddEndPuncttrue}
  {\def\EndOfBibitem{\unskip.}}
\providecommand*{\mciteBstWouldAddEndPunctfalse}
  {\let\EndOfBibitem\relax}
\providecommand*{\mciteSetBstMidEndSepPunct}[3]{}
\providecommand*{\mciteSetBstSublistLabelBeginEnd}[3]{}
\providecommand*{\EndOfBibitem}{}
\mciteSetBstSublistMode{f}
\mciteSetBstMaxWidthForm{subitem}
{(\emph{\alph{mcitesubitemcount}})}
\mciteSetBstSublistLabelBeginEnd{\mcitemaxwidthsubitemform\space}
{\relax}{\relax}

\bibitem[Benedek and Fennie(2011)]{Benedek2011}
N.~A. Benedek and C.~J. Fennie, \emph{Phys. Rev. Lett.}, 2011, \textbf{106},
  107204\relax
\mciteBstWouldAddEndPuncttrue
\mciteSetBstMidEndSepPunct{\mcitedefaultmidpunct}
{\mcitedefaultendpunct}{\mcitedefaultseppunct}\relax
\EndOfBibitem
\bibitem[Pitcher \emph{et~al.}(2015)Pitcher, Mandal, Dyer, Alaria, Borisov,
  Niu, Claridge, and Rosseinsky]{Pitcher2015}
M.~J. Pitcher, P.~Mandal, M.~S. Dyer, J.~Alaria, P.~Borisov, H.~Niu, J.~B.
  Claridge and M.~J. Rosseinsky, \emph{Science}, 2015, \textbf{347},
  420--424\relax
\mciteBstWouldAddEndPuncttrue
\mciteSetBstMidEndSepPunct{\mcitedefaultmidpunct}
{\mcitedefaultendpunct}{\mcitedefaultseppunct}\relax
\EndOfBibitem
\bibitem[Benedek \emph{et~al.}(2015)Benedek, Rondinelli, Djani, Ghosez, and
  Lightfoot]{Benedek2015}
N.~A. Benedek, J.~M. Rondinelli, H.~Djani, P.~Ghosez and P.~Lightfoot,
  \emph{Dalton Trans.}, 2015, \textbf{44}, 10543--10558\relax
\mciteBstWouldAddEndPuncttrue
\mciteSetBstMidEndSepPunct{\mcitedefaultmidpunct}
{\mcitedefaultendpunct}{\mcitedefaultseppunct}\relax
\EndOfBibitem
\bibitem[Senn \emph{et~al.}(2016)Senn, Murray, Luo, Wang, Huang, Cheong,
  Bombardi, Ablitt, Mostofi, and Bristowe]{Senn2016a}
M.~S. Senn, C.~A. Murray, X.~Luo, L.~Wang, F.-T. Huang, S.-W. Cheong,
  A.~Bombardi, C.~Ablitt, A.~A. Mostofi and N.~C. Bristowe, \emph{J. Am. Chem.
  Soc.}, 2016, \textbf{138}, 5479--5482\relax
\mciteBstWouldAddEndPuncttrue
\mciteSetBstMidEndSepPunct{\mcitedefaultmidpunct}
{\mcitedefaultendpunct}{\mcitedefaultseppunct}\relax
\EndOfBibitem
\bibitem[Glazer(1972)]{Glazer1972}
A.~M. Glazer, \emph{Acta Crystallogr. B}, 1972, \textbf{28}, 3384--3392\relax
\mciteBstWouldAddEndPuncttrue
\mciteSetBstMidEndSepPunct{\mcitedefaultmidpunct}
{\mcitedefaultendpunct}{\mcitedefaultseppunct}\relax
\EndOfBibitem
\bibitem[Howard and Stokes(1998)]{Howard1998}
C.~J. Howard and H.~T. Stokes, \emph{Acta Crystallogr. B}, 1998, \textbf{54},
  782--789\relax
\mciteBstWouldAddEndPuncttrue
\mciteSetBstMidEndSepPunct{\mcitedefaultmidpunct}
{\mcitedefaultendpunct}{\mcitedefaultseppunct}\relax
\EndOfBibitem
\bibitem[Howard \emph{et~al.}(2003)Howard, Kennedy, and Woodward]{Howard2003}
C.~J. Howard, B.~J. Kennedy and P.~M. Woodward, \emph{Acta Crystallogr. B},
  2003, \textbf{59}, 463--471\relax
\mciteBstWouldAddEndPuncttrue
\mciteSetBstMidEndSepPunct{\mcitedefaultmidpunct}
{\mcitedefaultendpunct}{\mcitedefaultseppunct}\relax
\EndOfBibitem
\bibitem[Dove(1993)]{Dove1993}
M.~T. Dove, \emph{{Introduction to Lattice Dynamics}}, Cambridge University
  Press, Cambridge, 1993\relax
\mciteBstWouldAddEndPuncttrue
\mciteSetBstMidEndSepPunct{\mcitedefaultmidpunct}
{\mcitedefaultendpunct}{\mcitedefaultseppunct}\relax
\EndOfBibitem
\bibitem[Snaith(2013)]{Snaith2013}
H.~J. Snaith, \emph{J. Phys. Chem. Lett.}, 2013, \textbf{4}, 3623--3630\relax
\mciteBstWouldAddEndPuncttrue
\mciteSetBstMidEndSepPunct{\mcitedefaultmidpunct}
{\mcitedefaultendpunct}{\mcitedefaultseppunct}\relax
\EndOfBibitem
\bibitem[Mitzi(2001)]{Mitzi2001}
D.~B. Mitzi, \emph{J. Chem. Soc. Dalt. Trans.}, 2001,  1--12\relax
\mciteBstWouldAddEndPuncttrue
\mciteSetBstMidEndSepPunct{\mcitedefaultmidpunct}
{\mcitedefaultendpunct}{\mcitedefaultseppunct}\relax
\EndOfBibitem
\bibitem[Tong \emph{et~al.}(2003)Tong, Ru, Wu, Chen, Chang, Mochizuki, and
  Kitagawa]{Tong2003}
M.-L. Tong, J.~Ru, Y.-M. Wu, X.-M. Chen, H.-C. Chang, K.~Mochizuki and
  S.~Kitagawa, \emph{New J. Chem.}, 2003, \textbf{27}, 779--782\relax
\mciteBstWouldAddEndPuncttrue
\mciteSetBstMidEndSepPunct{\mcitedefaultmidpunct}
{\mcitedefaultendpunct}{\mcitedefaultseppunct}\relax
\EndOfBibitem
\bibitem[Schlueter \emph{et~al.}(2004)Schlueter, Manson, Hyzer, and
  Geiser]{Schlueter2004}
J.~A. Schlueter, J.~L. Manson, K.~A. Hyzer and U.~Geiser, \emph{Inorg. Chem.},
  2004, \textbf{43}, 4100--4102\relax
\mciteBstWouldAddEndPuncttrue
\mciteSetBstMidEndSepPunct{\mcitedefaultmidpunct}
{\mcitedefaultendpunct}{\mcitedefaultseppunct}\relax
\EndOfBibitem
\bibitem[Schlueter \emph{et~al.}(2005)Schlueter, Manson, and
  Geiser]{Schlueter2005}
J.~A. Schlueter, J.~L. Manson and U.~Geiser, \emph{Inorg. Chem.}, 2005,
  \textbf{44}, 3194--3202\relax
\mciteBstWouldAddEndPuncttrue
\mciteSetBstMidEndSepPunct{\mcitedefaultmidpunct}
{\mcitedefaultendpunct}{\mcitedefaultseppunct}\relax
\EndOfBibitem
\bibitem[Berm{\'{u}}dez-Garc{\'{i}}a
  \emph{et~al.}(2016)Berm{\'{u}}dez-Garc{\'{i}}a, S{\'{a}}nchez-And{\'{u}}jar,
  Y{\'{a}}{\~{n}}ez-Vilar, Castro-Garc{\'{i}}a, Artiaga, L{\'{o}}pez-Beceiro,
  Botana, Alegria, and Se{\~{n}}aris-Rodriguez]{Bermudez-Garcia2016}
J.~M. Berm{\'{u}}dez-Garc{\'{i}}a, M.~S{\'{a}}nchez-And{\'{u}}jar,
  S.~Y{\'{a}}{\~{n}}ez-Vilar, S.~Castro-Garc{\'{i}}a, R.~Artiaga,
  J.~L{\'{o}}pez-Beceiro, L.~Botana, A.~Alegria and M.~A.
  Se{\~{n}}aris-Rodriguez, \emph{J. Mater. Chem. C}, 2016, \textbf{4},
  4889--4898\relax
\mciteBstWouldAddEndPuncttrue
\mciteSetBstMidEndSepPunct{\mcitedefaultmidpunct}
{\mcitedefaultendpunct}{\mcitedefaultseppunct}\relax
\EndOfBibitem
\bibitem[Du \emph{et~al.}(2015)Du, Xu, Huang, Su, Xue, He, Zhang, and
  Chen]{Du2015a}
Z.-Y. Du, T.-T. Xu, B.~Huang, Y.-J. Su, W.~Xue, C.-T. He, W.-X. Zhang and X.-M.
  Chen, \emph{Angew. Chem. Int. Ed.}, 2015, \textbf{54}, 914--918\relax
\mciteBstWouldAddEndPuncttrue
\mciteSetBstMidEndSepPunct{\mcitedefaultmidpunct}
{\mcitedefaultendpunct}{\mcitedefaultseppunct}\relax
\EndOfBibitem
\bibitem[G{\'{o}}mez-Aguirre \emph{et~al.}(2016)G{\'{o}}mez-Aguirre,
  Pato-Dold{\'{a}}n, Stroppa, Yang, Frauenheim, Mira, Y{\'{a}}{\~{n}}ez-Vilar,
  Artiaga, Castro-Garc{\'{i}}a, S{\'{a}}nchez-And{\'{u}}jar, and
  Se{\~{n}}ar{\'{i}}s-Rodr{\'{i}}guez]{Gomez-Aguirre2016}
L.~C. G{\'{o}}mez-Aguirre, B.~Pato-Dold{\'{a}}n, A.~Stroppa, L.-M. Yang,
  T.~Frauenheim, J.~Mira, S.~Y{\'{a}}{\~{n}}ez-Vilar, R.~Artiaga,
  S.~Castro-Garc{\'{i}}a, M.~S{\'{a}}nchez-And{\'{u}}jar and M.~A.
  Se{\~{n}}ar{\'{i}}s-Rodr{\'{i}}guez, \emph{Chem. Eur. J.}, 2016, \textbf{22},
  7863 --7870\relax
\mciteBstWouldAddEndPuncttrue
\mciteSetBstMidEndSepPunct{\mcitedefaultmidpunct}
{\mcitedefaultendpunct}{\mcitedefaultseppunct}\relax
\EndOfBibitem
\bibitem[Buser \emph{et~al.}(1977)Buser, Schwarzenbach, Petter, and
  Ludi]{Buser1977}
H.~J. Buser, D.~Schwarzenbach, W.~Petter and A.~Ludi, \emph{Inorg. Chem.},
  1977, \textbf{16}, 2704--2710\relax
\mciteBstWouldAddEndPuncttrue
\mciteSetBstMidEndSepPunct{\mcitedefaultmidpunct}
{\mcitedefaultendpunct}{\mcitedefaultseppunct}\relax
\EndOfBibitem
\bibitem[Aguil{\`{a}} \emph{et~al.}(2016)Aguil{\`{a}}, Prado, Koumousi,
  Mathoni{\`{e}}re, and Cl{\'{e}}rac]{Aguila2016}
D.~Aguil{\`{a}}, Y.~Prado, E.~S. Koumousi, C.~Mathoni{\`{e}}re and
  R.~Cl{\'{e}}rac, \emph{Chem. Soc. Rev.}, 2016, \textbf{45}, 203--224\relax
\mciteBstWouldAddEndPuncttrue
\mciteSetBstMidEndSepPunct{\mcitedefaultmidpunct}
{\mcitedefaultendpunct}{\mcitedefaultseppunct}\relax
\EndOfBibitem
\bibitem[Lefebvre \emph{et~al.}(2007)Lefebvre, Chartrand, and
  Leznoff]{Lefebvre2007}
J.~Lefebvre, D.~Chartrand and D.~B. Leznoff, \emph{Polyhedron}, 2007,
  \textbf{26}, 2189--2199\relax
\mciteBstWouldAddEndPuncttrue
\mciteSetBstMidEndSepPunct{\mcitedefaultmidpunct}
{\mcitedefaultendpunct}{\mcitedefaultseppunct}\relax
\EndOfBibitem
\bibitem[Hill \emph{et~al.}(2016)Hill, Thompson, and Goodwin]{Hill2016}
J.~A. Hill, A.~L. Thompson and A.~L. Goodwin, \emph{J. Am. Chem. Soc.}, 2016,
  \textbf{138}, 5886--5896\relax
\mciteBstWouldAddEndPuncttrue
\mciteSetBstMidEndSepPunct{\mcitedefaultmidpunct}
{\mcitedefaultendpunct}{\mcitedefaultseppunct}\relax
\EndOfBibitem
\bibitem[Sletten and Jensen(1973)]{Sletten1973}
E.~Sletten and L.~H. Jensen, \emph{Acta Crystallogr. B}, 1973, \textbf{29},
  1752--1756\relax
\mciteBstWouldAddEndPuncttrue
\mciteSetBstMidEndSepPunct{\mcitedefaultmidpunct}
{\mcitedefaultendpunct}{\mcitedefaultseppunct}\relax
\EndOfBibitem
\bibitem[Wang \emph{et~al.}(2004)Wang, Zhang, Otsuka, Inoue, Kobayashi, and
  Kurmoo]{Wang2004a}
Z.~Wang, B.~Zhang, T.~Otsuka, K.~Inoue, H.~Kobayashi and M.~Kurmoo,
  \emph{Dalton Trans.}, 2004,  2209--2216\relax
\mciteBstWouldAddEndPuncttrue
\mciteSetBstMidEndSepPunct{\mcitedefaultmidpunct}
{\mcitedefaultendpunct}{\mcitedefaultseppunct}\relax
\EndOfBibitem
\bibitem[Wang \emph{et~al.}(2004)Wang, Gan, Zhang, and Gao]{Wang2004}
X.-Y. Wang, L.~Gan, S.-W. Zhang and S.~Gao, \emph{Inorg. Chem.}, 2004,
  \textbf{43}, 4615--4625\relax
\mciteBstWouldAddEndPuncttrue
\mciteSetBstMidEndSepPunct{\mcitedefaultmidpunct}
{\mcitedefaultendpunct}{\mcitedefaultseppunct}\relax
\EndOfBibitem
\bibitem[Du \emph{et~al.}(2014)Du, Zhao, He, Wang, Xue, Zhou, Bai, Huang,
  Zhang, and Chen]{Du2014}
Z.-Y. Du, Y.-P. Zhao, C.-T. He, B.-Y. Wang, W.~Xue, H.-L. Zhou, J.~Bai,
  B.~Huang, W.-X. Zhang and X.-M. Chen, \emph{Cryst. Growth Des.}, 2014,
  \textbf{14}, 3903--3909\relax
\mciteBstWouldAddEndPuncttrue
\mciteSetBstMidEndSepPunct{\mcitedefaultmidpunct}
{\mcitedefaultendpunct}{\mcitedefaultseppunct}\relax
\EndOfBibitem
\bibitem[Duyker \emph{et~al.}(2016)Duyker, Hill, Howard, and
  Goodwin]{Duyker2016}
S.~G. Duyker, J.~A. Hill, C.~J. Howard and A.~L. Goodwin, \emph{J. Am. Chem.
  Soc.}, 2016,  DOI: 10.1021/jacs.6b06785\relax
\mciteBstWouldAddEndPuncttrue
\mciteSetBstMidEndSepPunct{\mcitedefaultmidpunct}
{\mcitedefaultendpunct}{\mcitedefaultseppunct}\relax
\EndOfBibitem
\bibitem[Kareis \emph{et~al.}(2012)Kareis, Lapidus, Her, Stephens, and
  Miller]{Kareis2012}
C.~M. Kareis, S.~H. Lapidus, J.-H. Her, P.~W. Stephens and J.~S. Miller,
  \emph{J. Am. Chem. Soc.}, 2012, \textbf{134}, 2246--2254\relax
\mciteBstWouldAddEndPuncttrue
\mciteSetBstMidEndSepPunct{\mcitedefaultmidpunct}
{\mcitedefaultendpunct}{\mcitedefaultseppunct}\relax
\EndOfBibitem
\bibitem[Evans \emph{et~al.}(2016)Evans, Thygesen, Bostr{\"{o}}m, Reynolds,
  Collings, Phillips, and Goodwin]{Evans2016}
N.~L. Evans, P.~M.~M. Thygesen, H.~L.~B. Bostr{\"{o}}m, E.~M. Reynolds, I.~E.
  Collings, A.~E. Phillips and A.~L. Goodwin, \emph{J. Am. Chem. Soc.}, 2016,
  \textbf{138}, 9393--9396\relax
\mciteBstWouldAddEndPuncttrue
\mciteSetBstMidEndSepPunct{\mcitedefaultmidpunct}
{\mcitedefaultendpunct}{\mcitedefaultseppunct}\relax
\EndOfBibitem
\bibitem[Zhang \emph{et~al.}(2015)Zhang, Shao, Li, Cai, Yao, Xiong, and
  Zhang]{Zhang2015}
X.~Zhang, X.-D. Shao, S.-C. Li, Y.~Cai, Y.-F. Yao, R.-G. Xiong and W.~Zhang,
  \emph{Chem. Commun.}, 2015, \textbf{51}, 4568--4571\relax
\mciteBstWouldAddEndPuncttrue
\mciteSetBstMidEndSepPunct{\mcitedefaultmidpunct}
{\mcitedefaultendpunct}{\mcitedefaultseppunct}\relax
\EndOfBibitem
\bibitem[Xu \emph{et~al.}(2016)Xu, Du, Zhang, and Chen]{Xu2016a}
W.-J. Xu, Z.-Y. Du, W.-X. Zhang and X.-M. Chen, \emph{CrystEngComm}, 2016\relax
\mciteBstWouldAddEndPuncttrue
\mciteSetBstMidEndSepPunct{\mcitedefaultmidpunct}
{\mcitedefaultendpunct}{\mcitedefaultseppunct}\relax
\EndOfBibitem
\bibitem[Stroppa \emph{et~al.}(2013)Stroppa, Barone, Jain, Perez-Mato, and
  Picozzi]{Stroppa2013}
A.~Stroppa, P.~Barone, P.~Jain, J.~M. Perez-Mato and S.~Picozzi, \emph{Adv.
  Mater.}, 2013, \textbf{25}, 2284--2290\relax
\mciteBstWouldAddEndPuncttrue
\mciteSetBstMidEndSepPunct{\mcitedefaultmidpunct}
{\mcitedefaultendpunct}{\mcitedefaultseppunct}\relax
\EndOfBibitem
\bibitem[Giddy \emph{et~al.}(1993)Giddy, Dove, Pawley, and Heine]{Giddy1993}
A.~P. Giddy, M.~T. Dove, G.~S. Pawley and V.~Heine, \emph{Acta Crystallogr. A},
  1993, \textbf{49}, 697--703\relax
\mciteBstWouldAddEndPuncttrue
\mciteSetBstMidEndSepPunct{\mcitedefaultmidpunct}
{\mcitedefaultendpunct}{\mcitedefaultseppunct}\relax
\EndOfBibitem
\bibitem[Goodwin(2006)]{Goodwin2006}
A.~L. Goodwin, \emph{Phys. Rev. B}, 2006, \textbf{74}, 134302\relax
\mciteBstWouldAddEndPuncttrue
\mciteSetBstMidEndSepPunct{\mcitedefaultmidpunct}
{\mcitedefaultendpunct}{\mcitedefaultseppunct}\relax
\EndOfBibitem
\bibitem[Pawley(1972)]{Pawley1972}
G.~S. Pawley, \emph{Phys. status solidi}, 1972, \textbf{49}, 475--488\relax
\mciteBstWouldAddEndPuncttrue
\mciteSetBstMidEndSepPunct{\mcitedefaultmidpunct}
{\mcitedefaultendpunct}{\mcitedefaultseppunct}\relax
\EndOfBibitem
\bibitem[Nanthamathee \emph{et~al.}(2015)Nanthamathee, Ling, Slater, and
  Attfield]{Nanthamathee2015}
C.~Nanthamathee, S.~Ling, B.~Slater and M.~P. Attfield, \emph{Chem. Mater.},
  2015, \textbf{27}, 85--95\relax
\mciteBstWouldAddEndPuncttrue
\mciteSetBstMidEndSepPunct{\mcitedefaultmidpunct}
{\mcitedefaultendpunct}{\mcitedefaultseppunct}\relax
\EndOfBibitem
\bibitem[Serra-Crespo \emph{et~al.}(2015)Serra-Crespo, Dikhtiarenko, Stavitski,
  Juan-Alca{\~{n}}iz, Kapteijn, Coudert, and Gascon]{Serra-Crespo2015}
P.~Serra-Crespo, A.~Dikhtiarenko, E.~Stavitski, J.~Juan-Alca{\~{n}}iz,
  F.~Kapteijn, F.-X. Coudert and J.~Gascon, \emph{CrystEngComm}, 2015,
  \textbf{17}, 276--280\relax
\mciteBstWouldAddEndPuncttrue
\mciteSetBstMidEndSepPunct{\mcitedefaultmidpunct}
{\mcitedefaultendpunct}{\mcitedefaultseppunct}\relax
\EndOfBibitem
\bibitem[Aizu(1969)]{Aizu1969}
K.~Aizu, \emph{J. Phys. Soc. Jpn.}, 1969, \textbf{27}, 1171--1178\relax
\mciteBstWouldAddEndPuncttrue
\mciteSetBstMidEndSepPunct{\mcitedefaultmidpunct}
{\mcitedefaultendpunct}{\mcitedefaultseppunct}\relax
\EndOfBibitem
\bibitem[Salje(2012)]{Salje2012a}
E.~K. Salje, \emph{Annu. Rev. Mater. Res.}, 2012, \textbf{42}, 265--283\relax
\mciteBstWouldAddEndPuncttrue
\mciteSetBstMidEndSepPunct{\mcitedefaultmidpunct}
{\mcitedefaultendpunct}{\mcitedefaultseppunct}\relax
\EndOfBibitem
\bibitem[Ortiz \emph{et~al.}(2012)Ortiz, Boutin, Fuchs, and Coudert]{Ortiz2012}
A.~U. Ortiz, A.~Boutin, A.~H. Fuchs and F.~X. Coudert, \emph{Phys. Rev. Lett.},
  2012, \textbf{109}, 1--5\relax
\mciteBstWouldAddEndPuncttrue
\mciteSetBstMidEndSepPunct{\mcitedefaultmidpunct}
{\mcitedefaultendpunct}{\mcitedefaultseppunct}\relax
\EndOfBibitem
\bibitem[Coudert(2015)]{Coudert2015}
F.~X. Coudert, \emph{Chem. Mater.}, 2015, \textbf{27}, 1905--1916\relax
\mciteBstWouldAddEndPuncttrue
\mciteSetBstMidEndSepPunct{\mcitedefaultmidpunct}
{\mcitedefaultendpunct}{\mcitedefaultseppunct}\relax
\EndOfBibitem
\bibitem[Hunt \emph{et~al.}(2015)Hunt, Cliffe, Hill, Cairns, Funnell, and
  Goodwin]{Hunt2015}
S.~J. Hunt, M.~J. Cliffe, J.~A. Hill, A.~B. Cairns, N.~P. Funnell and A.~L.
  Goodwin, \emph{CrystEngComm}, 2015, \textbf{17}, 361--369\relax
\mciteBstWouldAddEndPuncttrue
\mciteSetBstMidEndSepPunct{\mcitedefaultmidpunct}
{\mcitedefaultendpunct}{\mcitedefaultseppunct}\relax
\EndOfBibitem
\bibitem[Cairns and Goodwin(2015)]{Cairns2015}
A.~B. Cairns and A.~L. Goodwin, \emph{Phys. Chem. Chem. Phys.}, 2015,
  \textbf{17}, 20449--20465\relax
\mciteBstWouldAddEndPuncttrue
\mciteSetBstMidEndSepPunct{\mcitedefaultmidpunct}
{\mcitedefaultendpunct}{\mcitedefaultseppunct}\relax
\EndOfBibitem
\bibitem[Redfern and Salje(1988)]{Redfern1988}
S.~A.~T. Redfern and T.~E. Salje, \emph{J. Phys. C Solid State Phys.}, 1988,
  \textbf{21}, 277--285\relax
\mciteBstWouldAddEndPuncttrue
\mciteSetBstMidEndSepPunct{\mcitedefaultmidpunct}
{\mcitedefaultendpunct}{\mcitedefaultseppunct}\relax
\EndOfBibitem
\bibitem[Kennedy \emph{et~al.}(1999)Kennedy, Howard, and
  Chakoumakos]{Kennedy1999}
B.~J. Kennedy, C.~J. Howard and B.~C. Chakoumakos, \emph{J. Phys. Condens.
  Matter}, 1999, \textbf{11}, 1479--1488\relax
\mciteBstWouldAddEndPuncttrue
\mciteSetBstMidEndSepPunct{\mcitedefaultmidpunct}
{\mcitedefaultendpunct}{\mcitedefaultseppunct}\relax
\EndOfBibitem
\bibitem[Bradley and Cracknell(1972)]{Bradley1972}
G.~J. Bradley and A.~P. Cracknell, \emph{{The Mathematical Theory of Symmetry
  in Solids}}, Clarendon Press, Oxford, 1972\relax
\mciteBstWouldAddEndPuncttrue
\mciteSetBstMidEndSepPunct{\mcitedefaultmidpunct}
{\mcitedefaultendpunct}{\mcitedefaultseppunct}\relax
\EndOfBibitem
\bibitem[Mautner \emph{et~al.}(1988)Mautner, Krischner, and
  Kratky]{Mautner1988}
F.~A. Mautner, H.~Krischner and C.~Kratky, \emph{Monatshefte f{\"{u}}r Chemie},
  1988, \textbf{119}, 1245--1249\relax
\mciteBstWouldAddEndPuncttrue
\mciteSetBstMidEndSepPunct{\mcitedefaultmidpunct}
{\mcitedefaultendpunct}{\mcitedefaultseppunct}\relax
\EndOfBibitem
\bibitem[Campbell \emph{et~al.}(2006)Campbell, Stokes, Tanner, and
  Hatch]{Campbell2006}
B.~J. Campbell, H.~T. Stokes, D.~E. Tanner and D.~M. Hatch, \emph{J. Appl.
  Crystallogr.}, 2006, \textbf{39}, 607--614\relax
\mciteBstWouldAddEndPuncttrue
\mciteSetBstMidEndSepPunct{\mcitedefaultmidpunct}
{\mcitedefaultendpunct}{\mcitedefaultseppunct}\relax
\EndOfBibitem
\bibitem[Zhao \emph{et~al.}(2013)Zhao, Huang, Zhang, Shao, Wei, and
  Wang]{Zhao2013}
X.-H. Zhao, X.-C. Huang, S.-L. Zhang, D.~Shao, H.-Y. Wei and X.-Y. Wang,
  \emph{J. Am. Chem. Soc.}, 2013, \textbf{135}, 16006--16009\relax
\mciteBstWouldAddEndPuncttrue
\mciteSetBstMidEndSepPunct{\mcitedefaultmidpunct}
{\mcitedefaultendpunct}{\mcitedefaultseppunct}\relax
\EndOfBibitem
\bibitem[Peel \emph{et~al.}(2012)Peel, Thompson, Daoud-Aladine, Ashbrook, and
  Lightfoot]{Peel2012}
M.~D. Peel, S.~P. Thompson, A.~Daoud-Aladine, S.~E. Ashbrook and P.~Lightfoot,
  \emph{Inorg. Chem.}, 2012, \textbf{51}, 6876--6889\relax
\mciteBstWouldAddEndPuncttrue
\mciteSetBstMidEndSepPunct{\mcitedefaultmidpunct}
{\mcitedefaultendpunct}{\mcitedefaultseppunct}\relax
\EndOfBibitem
\bibitem[Cowley(1964)]{Cowley1964}
R.~A. Cowley, \emph{Phys. Rev.}, 1964, \textbf{134}, A981--A997\relax
\mciteBstWouldAddEndPuncttrue
\mciteSetBstMidEndSepPunct{\mcitedefaultmidpunct}
{\mcitedefaultendpunct}{\mcitedefaultseppunct}\relax
\EndOfBibitem
\bibitem[Shirane(1974)]{Shirane1974}
G.~Shirane, \emph{Rev. Mod. Phys.}, 1974, \textbf{46}, 437--449\relax
\mciteBstWouldAddEndPuncttrue
\mciteSetBstMidEndSepPunct{\mcitedefaultmidpunct}
{\mcitedefaultendpunct}{\mcitedefaultseppunct}\relax
\EndOfBibitem
\bibitem[Dove(2002)]{Dove2002}
M.~T. Dove, \emph{Eur. J. Miner.}, 2002, \textbf{14}, 203--224\relax
\mciteBstWouldAddEndPuncttrue
\mciteSetBstMidEndSepPunct{\mcitedefaultmidpunct}
{\mcitedefaultendpunct}{\mcitedefaultseppunct}\relax
\EndOfBibitem
\bibitem[Gr{\"{u}}neisen(1912)]{Gruneisen1912}
F.~Gr{\"{u}}neisen, \emph{Ann. Phys.}, 1912, \textbf{344}, 257--306\relax
\mciteBstWouldAddEndPuncttrue
\mciteSetBstMidEndSepPunct{\mcitedefaultmidpunct}
{\mcitedefaultendpunct}{\mcitedefaultseppunct}\relax
\EndOfBibitem
\bibitem[Gale(1997)]{Gale1997}
S.~D. Gale, \emph{J. Chem. Soc., Faraday Trans.}, 1997, \textbf{93},
  629--637\relax
\mciteBstWouldAddEndPuncttrue
\mciteSetBstMidEndSepPunct{\mcitedefaultmidpunct}
{\mcitedefaultendpunct}{\mcitedefaultseppunct}\relax
\EndOfBibitem
\bibitem[Evans(1999)]{Evans1999a}
J.~S.~O. Evans, \emph{J. Chem. Soc. Dalt. Trans.}, 1999, \textbf{3},
  3317--3326\relax
\mciteBstWouldAddEndPuncttrue
\mciteSetBstMidEndSepPunct{\mcitedefaultmidpunct}
{\mcitedefaultendpunct}{\mcitedefaultseppunct}\relax
\EndOfBibitem
\bibitem[Goodwin \emph{et~al.}(2005)Goodwin, Chapman, and Kepert]{Goodwin2005a}
A.~L. Goodwin, K.~W. Chapman and C.~J. Kepert, \emph{J. Am. Chem. Soc.}, 2005,
  \textbf{127}, 17980--17981\relax
\mciteBstWouldAddEndPuncttrue
\mciteSetBstMidEndSepPunct{\mcitedefaultmidpunct}
{\mcitedefaultendpunct}{\mcitedefaultseppunct}\relax
\EndOfBibitem
\bibitem[Chapman \emph{et~al.}(2006)Chapman, Chupas, and Kepert]{Chapman2006}
K.~W. Chapman, P.~J. Chupas and C.~J. Kepert, \emph{J. Am. Chem. Soc.}, 2006,
  \textbf{128}, 7009--7014\relax
\mciteBstWouldAddEndPuncttrue
\mciteSetBstMidEndSepPunct{\mcitedefaultmidpunct}
{\mcitedefaultendpunct}{\mcitedefaultseppunct}\relax
\EndOfBibitem
\bibitem[Adak \emph{et~al.}(2011)Adak, Daemen, Hartl, Williams, Summerhill, and
  Nakotte]{Adak2011}
S.~Adak, L.~L. Daemen, M.~Hartl, D.~Williams, J.~Summerhill and H.~Nakotte,
  \emph{J. Solid State Chem.}, 2011, \textbf{184}, 2854--2861\relax
\mciteBstWouldAddEndPuncttrue
\mciteSetBstMidEndSepPunct{\mcitedefaultmidpunct}
{\mcitedefaultendpunct}{\mcitedefaultseppunct}\relax
\EndOfBibitem
\bibitem[Rimmer \emph{et~al.}(2014)Rimmer, Dove, Goodwin, and
  Palmer]{Rimmer2014}
L.~H.~N. Rimmer, M.~T. Dove, A.~L. Goodwin and D.~C. Palmer, \emph{Phys. Chem.
  Chem. Phys.}, 2014, \textbf{16}, 21144--21152\relax
\mciteBstWouldAddEndPuncttrue
\mciteSetBstMidEndSepPunct{\mcitedefaultmidpunct}
{\mcitedefaultendpunct}{\mcitedefaultseppunct}\relax
\EndOfBibitem
\bibitem[Trousselet \emph{et~al.}(2015)Trousselet, Boutin, and
  Coudert]{Trousselet2015}
F.~Trousselet, A.~Boutin and F.-X. Coudert, \emph{Chem. Mater.}, 2015,
  \textbf{27}, 4422--4430\relax
\mciteBstWouldAddEndPuncttrue
\mciteSetBstMidEndSepPunct{\mcitedefaultmidpunct}
{\mcitedefaultendpunct}{\mcitedefaultseppunct}\relax
\EndOfBibitem
\bibitem[Fang \emph{et~al.}(2013)Fang, Dove, Rimmer, and Misquitta]{Fang2013}
H.~Fang, M.~T. Dove, L.~H.~N. Rimmer and A.~J. Misquitta, \emph{Phys. Rev. B},
  2013, \textbf{88}, 104306\relax
\mciteBstWouldAddEndPuncttrue
\mciteSetBstMidEndSepPunct{\mcitedefaultmidpunct}
{\mcitedefaultendpunct}{\mcitedefaultseppunct}\relax
\EndOfBibitem
\bibitem[Oh \emph{et~al.}(2015)Oh, Luo, Huang, Wang, and Cheong]{Oh2015}
Y.~S. Oh, X.~Luo, F.-T. Huang, Y.~Wang and S.-W. Cheong, \emph{Nat. Mater.},
  2015, \textbf{14}, 407--413\relax
\mciteBstWouldAddEndPuncttrue
\mciteSetBstMidEndSepPunct{\mcitedefaultmidpunct}
{\mcitedefaultendpunct}{\mcitedefaultseppunct}\relax
\EndOfBibitem
\bibitem[Stroppa \emph{et~al.}(2011)Stroppa, Jain, Barone, Marsman, Perez-Mato,
  Cheetham, Kroto, and Picozzi]{Stroppa2011}
A.~Stroppa, P.~Jain, P.~Barone, M.~Marsman, J.~M. Perez-Mato, A.~K. Cheetham,
  H.~W. Kroto and S.~Picozzi, \emph{Angew. Chem. Int. Ed.}, 2011, \textbf{50},
  5847--5850\relax
\mciteBstWouldAddEndPuncttrue
\mciteSetBstMidEndSepPunct{\mcitedefaultmidpunct}
{\mcitedefaultendpunct}{\mcitedefaultseppunct}\relax
\EndOfBibitem
\bibitem[Rondinelli and Fennie(2012)]{Rondinelli2012}
J.~M. Rondinelli and C.~J. Fennie, \emph{Adv. Mater.}, 2012, \textbf{24},
  1961--1968\relax
\mciteBstWouldAddEndPuncttrue
\mciteSetBstMidEndSepPunct{\mcitedefaultmidpunct}
{\mcitedefaultendpunct}{\mcitedefaultseppunct}\relax
\EndOfBibitem
\end{mcitethebibliography}
\bibliographystyle{rsc} 

\end{document}